\documentclass[sigconf]{acmart}
\AtBeginDocument{%
  }

\usepackage{booktabs}
\usepackage{multirow}
\usepackage{graphicx}
\usepackage{enumitem}
\usepackage{tabularx}
\usepackage[table,xcdraw]{xcolor}
\usepackage[most]{tcolorbox}

\newtheorem{lemma}{Lemma}

\newcommand{\ie}{\emph{i.e., }}
\newcommand{\eg}{\emph{e.g., }}

\newcommand{\cf}{\emph{cf. }}

\copyrightyear{2026}
\acmYear{2026}
\setcopyright{cc}
\setcctype{by}
\acmConference[WWW '26]{Proceedings of the ACM Web Conference 2026}{April 13--17, 2026}{Dubai, United Arab Emirates}
\acmBooktitle{Proceedings of the ACM Web Conference 2026 (WWW '26), April 13--17, 2026, Dubai, United Arab Emirates}
\acmPrice{}
\acmDOI{10.1145/3774904.3792607}
\acmISBN{979-8-4007-2307-0/2026/04}

\begin{document}

\title{Does LLM Focus on the Right Words? Mitigating Context Bias in LLM-based Recommenders}

\author{Bohao Wang}
\email{bohao.wang@zju.edu.cn}
\authornotemark[2]
\authornotemark[3]
\orcid{0009-0006-8264-3182}
\affiliation{%
  \institution{Zhejiang University}
  \city{Hangzhou}
  \country{China}
}

\author{Jiawei Chen}
\email{sleepyhunt@zju.edu.cn}
\orcid{0000-0002-4752-2629}
\authornote{Corresponding author.}
\authornote{State Key Laboratory of Blockchain and Data Security, Zhejiang University.}
\authornote{College of Computer Science and Technology, Zhejiang University.}
\authornote{Hangzhou High-Tech Zone (Binjiang) Institute of Blockchain and Data Security.}
\affiliation{%
  \institution{Zhejiang University}
  \city{Hangzhou}
  \country{China}
}

\author{Feng Liu}
\email{liufeng4hit@gmail.com}
\affiliation{%
  \institution{OPPO Research Institute}
  \city{Shenzhen}
  \country{China}
}

\author{Changwang Zhang}
\email{changwangzhang@foxmail.com}
\affiliation{%
  \institution{OPPO Research Institute}
  \city{Shenzhen}
  \country{China}
}

\author{Jun Wang}
\email{junwang.lu@gmail.com}
\affiliation{%
  \institution{OPPO Research Institute}
  \city{Shenzhen}
  \country{China}
}

\author{Canghong Jin}
\email{jinch@zucc.edu.cn}
\affiliation{%
  \institution{Hangzhou City University}
  \city{Hangzhou}
  \country{China}
}

\author{Chun Chen}
\email{chenc@cs.zju.edu.cn}
\authornotemark[2]
\authornotemark[3]
\orcid{0000-0002-6198-7481}
\affiliation{%
  \institution{Zhejiang University}
  \city{Hangzhou}
  \country{China}
}

\author{Can Wang}
\email{wcan@zju.edu.cn}
\authornotemark[2]
\authornotemark[4]
\orcid{0000-0002-5890-4307}
\affiliation{%
  \institution{Zhejiang University}
  \city{Hangzhou}
  \country{China}
}

\renewcommand{\shortauthors}{Bohao Wang et al.}



\begin{abstract}
Large language models (LLMs), owing to their extensive open-domain knowledge and semantic reasoning capabilities, have been increasingly integrated into recommender systems (RS). However, a substantial gap remains between the pre-training objectives of LLMs and the specific requirements of recommendation tasks. To address this gap, supervised fine-tuning (SFT) is commonly performed on specially curated recommendation datasets to further enhance their predictive ability. Despite its success, SFT exhibits a critical limitation: it induces \textbf{Context Bias}, whereby the model over-relies on auxiliary tokens—such as task descriptions and prefix-generated tokens—while underutilizing core user interaction tokens that encode user-specific preferences. This bias not only undermines recommendation accuracy but also raises unfairness concerns.

To address this issue, we propose \textbf{Group Distributionally Robust Optimization-based Tuning (GDRT)}, a novel fine-tuning paradigm that enforces consistent model performance across token groups with varying degrees of relevance to auxiliary tokens. By adaptively upweighting underperforming groups, typically those weakly correlated with auxiliary tokens, GDRT shifts the model's attention from superficial auxiliary cues to informative user interaction tokens, thereby mitigating context bias. Extensive experiments conducted on three public datasets demonstrate that GDRT effectively mitigates context bias, yielding substantial improvements in recommendation accuracy (with an average NDCG@10 gain of 24.29\%) and significantly enhancing recommendation fairness. The code is available at \url{https://github.com/WANGBohaO-jpg/GDRT}.

\end{abstract}



\maketitle

\section{Introduction}

\begin{figure*}[t]
  \centering
  \includegraphics[width=\linewidth]{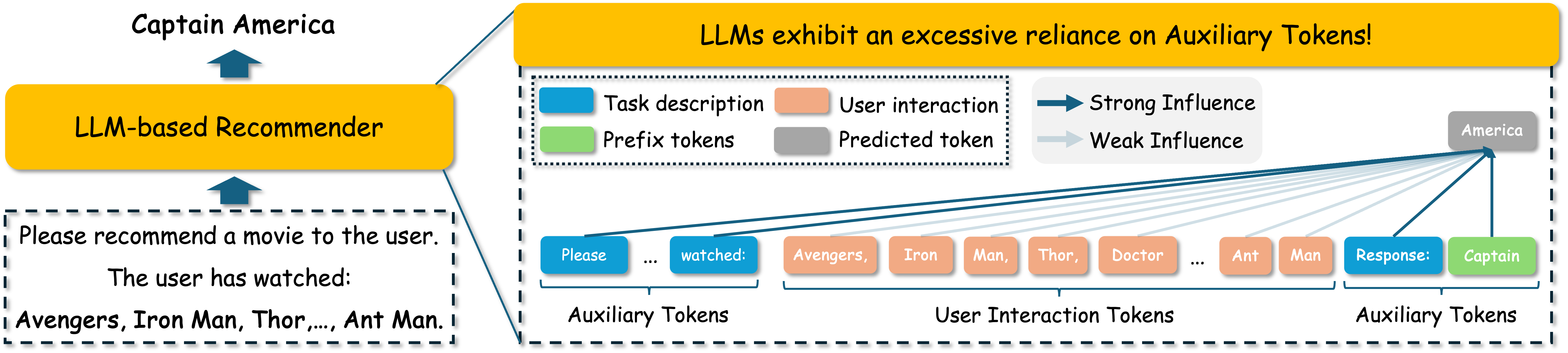}
  \caption{Illustration of LLM‑based recommendations and context bias, wherein the model exhibits an over‑reliance on auxiliary tokens (\ie \textcolor[HTML]{0e9ed5}{Task description}, \textcolor[HTML]{8ed973}{Prefix tokens}) and insufficient utilization of \textcolor[HTML]{f2aa84}{User interaction} during generation.}
  \Description{}
  \label{fig:bias_diagram}
\end{figure*}

\begin{figure}[t]
  \centering
  \includegraphics[width=\linewidth]{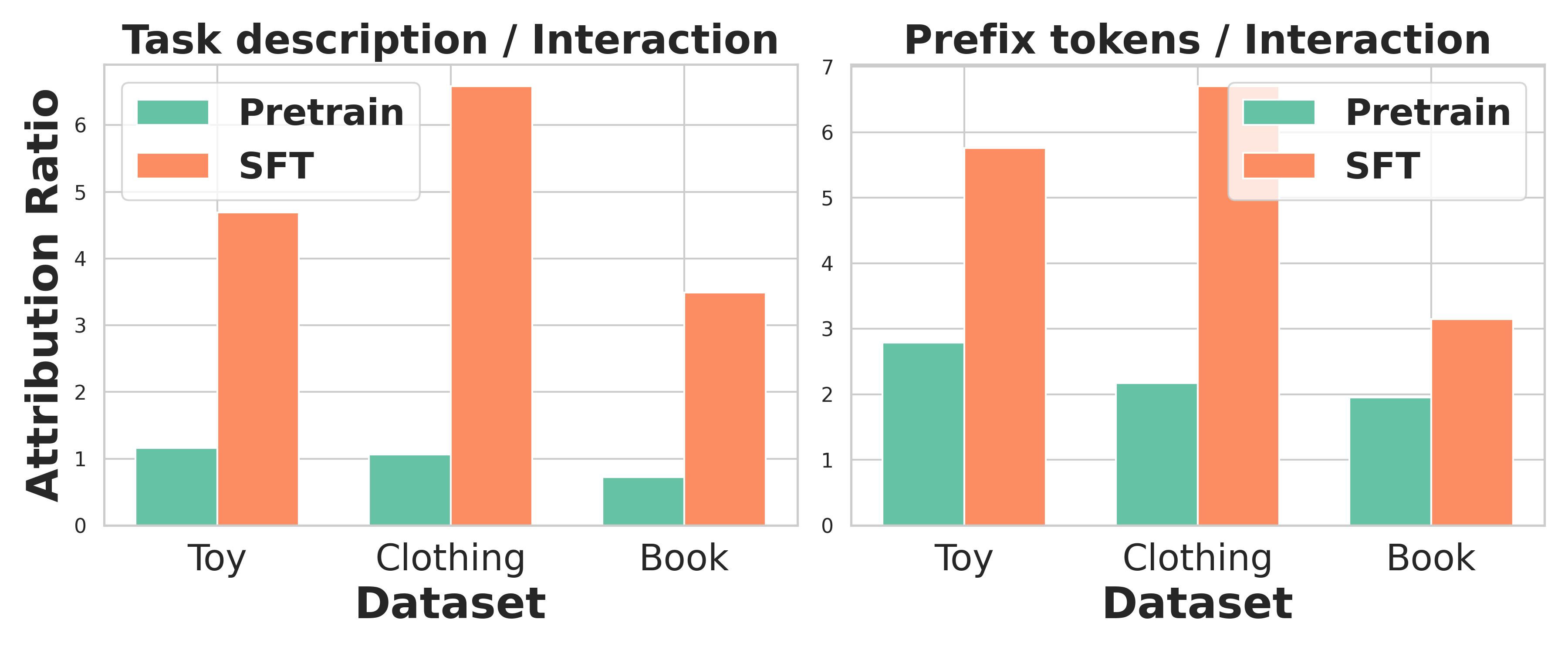}
  \caption{Ratio of attribution values between auxiliary tokens and user interaction tokens before and after SFT. Left: task description vs. user interaction tokens. Right: prefix tokens of predicted item (take the first token) vs. user interaction tokens.}
  \Description{}
  \label{fig:feature_ablation_intro}
\end{figure}

With remarkable open-domain knowledge and semantic reasoning capabilities \cite{guo2025deepseek, achiam2023gpt}, Large Language Models (LLMs) have been extensively explored for integration into recommendation systems (RS) \cite{wu2024survey}. One prominent approach involves positioning LLMs as the central recommendation backbone \cite{wang2025msl, li2023large, bao2025bi, bao2023tallrec, zhu2024collaborative, wang2023zero, tan2024idgenrec, zheng2024harnessing, wang2024flip, liao2024llara, kim2024large}. These methods express items as textual descriptions (\eg titles), and construct language prompts based on users' past interactions, which are then used to instruct LLMs to predict users' future interactions. Figure \ref{fig:bias_diagram} illustrates the mechanism of such LLM-based recommendation. LLMs operate at a fine-grained semantic token level, sequentially generating tokens of the predicted items by analyzing their nuanced semantic relations with the user's previous interactions and other auxiliary information (\eg task descriptions, prefix tokens of the predicted item). This fine-grained paradigm enables the capture of subtle semantic patterns in user interests and thus represents a promising direction for advancing recommender systems \cite{bao2025bi}.

To better align LLMs with recommendation objectives and capture collaborative filtering signals, Supervised Fine-Tuning (SFT) is commonly employed \cite{bao2025bi, bao2023tallrec, liao2024llara, wang2025msl}. In this strategy, each prompt is paired with the target item description, and the LLM is fine-tuned to generate the correct prediction. This procedure enables the model to capture semantic relationships between prompts and targets present in the training data, often resulting in substantial performance gains.

However, we find that SFT introduces a significant \textbf{Context Bias}. Specifically, SFT drives the model to over-rely on auxiliary tokens (\eg task descriptions or prefix tokens) while under-utilizing core interaction tokens that encode user personalized preferences as shown in Figure \ref{fig:bias_diagram}. To verify this, we conduct Feature Ablation Attribution analysis \cite{kokhlikyan2020captum, miglani2023using}, a standard approach to quantify each token’s contribution to model predictions. Figure \ref{fig:feature_ablation_intro} shows that SFT dramatically amplifies the relative impact of auxiliary tokens while significantly suppressing the influence of interaction tokens, with the influence ratio shifting from about 1:1 before fine-tuning to more than 6:1 afterward as measured on typical Amazon datasets.

\begin{figure}[t]
  \centering
  \includegraphics[width=\linewidth]{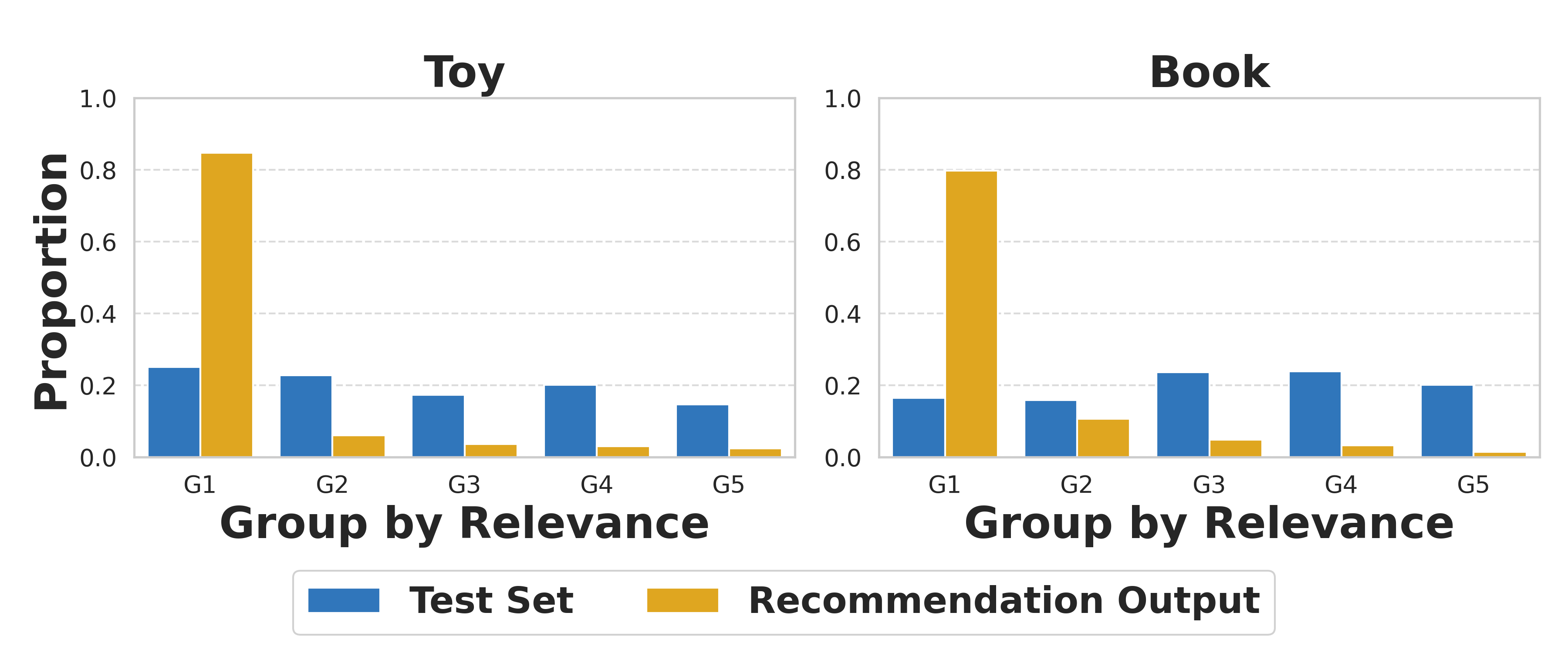}
  \caption{Distribution of Top‑1 recommended items generated by the SFT-trained model across five group defined according to the semantic relevance of items to the auxiliary tokens (Group 1: highest relevance, Group 5: lowest relevance). We also present the distribution of the test set across the same five groups for comparison.}
  \Description{}
  \label{fig:fairness_intro}
\end{figure}

This over-reliance reveals shortcut learning: LLMs simply memorize correlations with frequently occurring auxiliary tokens rather than grounding predictions in user-specific preferences. Such bias not only undermines recommendation accuracy but also raises serious fairness concerns. Specifically, this bias skews recommendations toward a narrow subset of items whose tokens exhibit higher semantic relevance to auxiliary tokens. To verify this, we divided items into five groups based on their relevance to auxiliary tokens. As shown in Figure \ref{fig:fairness_intro}, over 80\% of recommended items fall into the highest relevance group, while this group comprises only 20\% of target items in the test set. These findings motivate our core research question: \textbf{How can we mitigate context bias in LLM-based recommenders?}

To address this, we propose a novel fine-tuning strategy \textbf{GDRT}, that leverages Group Distributionally Robust Optimization (Group DRO) \cite{sagawa2019distributionally} to mitigate context bias. We first group training instances according to the semantic relevance between the target token and the auxiliary tokens, which can be evaluated by the LLM's predictive probability when user history is masked. Group DRO then enforces LLMs to perform consistently well across all groups, regardless of their relevance strength with auxiliary tokens. This optimization objective naturally shifts the LLM's attention away from auxiliary tokens toward user-specific interaction tokens, as simple reliance on shortcut auxiliary tokens results in poor performance on groups with weaker correlations. 
Importantly, GDRT is easy to implement and computationally efficient, requiring only high‑efficiency group construction and dynamic sample weighting during training. It can be seamlessly integrated into various LLM-based recommenders, yielding improvements in both accuracy and fairness.

Our main contributions are:
\begin{itemize}[left=5pt]
\item We provide a comprehensive empirical analysis revealing that SFT in LLM-based recommendation induces significant context bias, negatively affecting both accuracy and fairness.
\item We propose GDRT, a Group DRO-based fine-tuning strategy, to effectively mitigate context bias in LLM-based recommendation.
\item We conduct extensive experiments demonstrating that GDRT achieves state-of-the-art recommendation performance in both accuracy and fairness metrics.
\end{itemize}

\section{Preliminary}
\subsection{LLM-based Recommendation}
Following previous work \cite{liao2024llara, bao2025bi, na2024enhancing, bao2023tallrec, lin2024rella, zheng2024adapting}, this paper also focuses on sequential recommendation \cite{kang2018self}, a conventional recommendation scenario in practice.  Let \( \mathcal{V} \) denote the set of items in the recommendation system. Given a user's historical interaction sequence \( S = \{s_{1}, s_{2}, ..., s_{n}\} \), where each \( s_{i} \in \mathcal{V} \) represents the \( i \)-th interacted item, the goal of the RS is to predict the user's next interaction $s_{n+1}$ that the user is likely to interact with.

The remarkable success of LLMs across diverse domains \cite{wu2023visual, ouyang2022training, tang2024graphgpt, Chen_2025_ICCV, wang2025adaptive, chen2025arrows} has spurred growing interest in their application to recommendation systems (RS) \cite{wu2024survey}. A prominent approach is to directly leverage LLMs as recommenders \cite{li2023large}. As shown in Figure \ref{fig:bias_diagram}, this paradigm constructs a language prompt \(x = [\,x^{\text{task}};\, x^{\text{user}}\,]\), where $x^{\text{task}}$ represents the task description and $x^{\text{user}}$ denotes the textual form (\eg titles) of a user’s historical interactions. This prompt then guides the LLM to generate the descriptions of recommended items $y$. Notably, LLMs operate at a fine-grained semantic token level, sequentially generating tokens of the predicted items according to the model estimated probability $P_\theta(y_t|x,y_{<t})$, where $y_{t}$ denotes the $t$-th predictive tokens and $y_{<t}$ denotes the prefix tokens of the prediction. This fine-grained token-level paradigm has the potential to capture subtle semantic patterns in user preferences.

To align LLMs with recommendation objectives, supervised fine-tuning (SFT) is commonly applied, fine-tuning all or part of the model parameters using recommendation data \cite{bao2025bi, bao2023tallrec, liao2024llara, zhu2024collaborative}. 
In this process, the training data is reorganized into a set of prompt–target pairs $\mathcal{D} = \{(x_i,y^*_i)\}_{i=1}^{N}$ where each \(x_i\) is the constructed prompt and \(y^*_i\) is the textual description of the target item. The LLM is optimized with the following log-likelihood objective:
\begin{equation}
\label{eq:SFT_loss}
\mathcal{L}_{SFT} = - \frac{1}{N} \sum_{i=1}^{N} \sum_{t=1}^{|y^*_i|} \log P_\theta\!\left(y^*_{i,t} \mid x_i, y^*_{i,<t}\right)
\end{equation}
where \(|y^*_i|\) denotes the token length of the target item description. SFT increases the generative probability of the target item, encouraging the LLM to capture the inherent token-level semantic correlations between each target token $y_{i,t}^*$ and the user interactions $x^\text{user}_i$, task descriptions $x^{\text{task}}$ and the prefix tokens $y^*_{i,<t}$. This process often yields substantial performance improvements \cite{bao2023tallrec}.

\subsection{Analyses on Context Bias}
\label{sec:analyses_language_bias}
In this section, we first identify the context bias in fine-tuning LLM for recommendation, followed by discussing its negative effect. We then analyze the underlying causes of this bias and discuss why existing methods can not effectively address this issue. \footnote{The experimental configuration in this section is consistent with our main experimental setup described in Section \ref{sec:implementation_details}. We also provide additional analyses on other Prompt templates and LLMs in Appendix \ref{apd:FAA}.}

\subsubsection{Empirical Evidence Demonstrating Context Bias}
\label{sec:evidence_bias}
We conduct Feature Ablation Attribution (FAA) \cite{kokhlikyan2020captum, miglani2023using} analysis to quantify contribution of different token types to the model predictions. 
FAA is an attribution method used to evaluate the importance of individual input components by measuring how the model’s output changes when specific inputs are masked, and it has been widely adopted in the LLMs for interpreting token-level contributions \cite{zhou2024explaining, zhao2024explainability, barkan2024llm}. 
A higher attribution value indicates a greater influence of input tokens on the model's output. Figure \ref{fig:feature_ablation_intro} shows the ratio of attribution values between auxiliary tokens (\eg task descriptions or prefix tokens) and user-specific interaction tokens. We report this ratio both before and after fine-tuning to enable direct comparison. 
From these results, we make the following observation:

\textit{\textbf{Context Bias:} Supervised fine-tuning (SFT) can bias LLMs to over-rely on auxiliary tokens while under-utilizing core interaction tokens that encode user personalized preferences.}

Before fine-tuning, the attribution value ratio between task description and interaction tokens is approximately 1:1. After fine-tuning, this ratio is markedly amplified across all datasets (\eg Toy: 4.69:1; Clothing: 6.58:1). A similar phenomenon is observed when comparing prefix tokens with interaction tokens. These results clearly indicate the presence of context bias, whereby the model disproportionately relies on auxiliary tokens rather than more informative interaction tokens.





\begin{figure}[t]
  \centering
  \includegraphics[width=\linewidth]{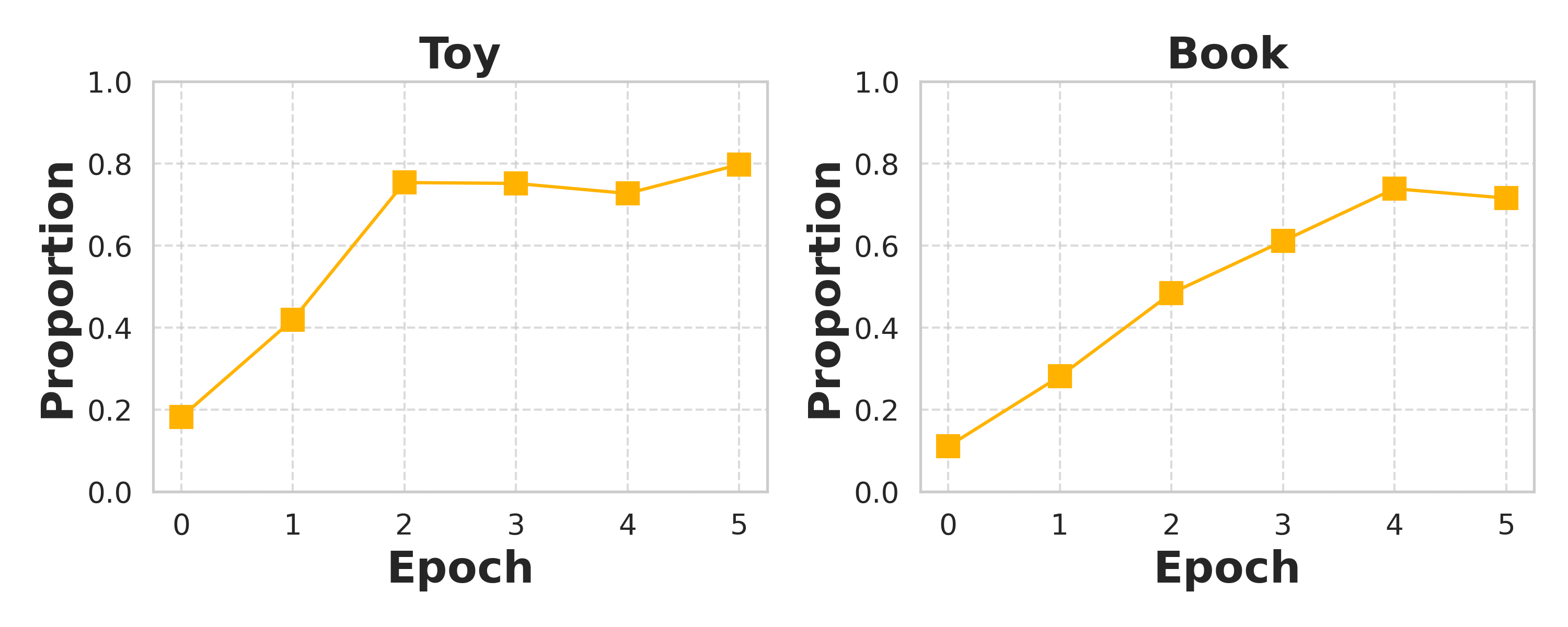}
  \caption{Proportion of Top‑1 recommendations belonging to Item Group 1 (highest relevance with auxiliary tokens) over the course of SFT.}
  \Description{}
  \label{fig:fairness_curve_intro}
\end{figure}

\subsubsection{Negative Effect of Context Bias}
Context bias can substantially hinder the effectiveness of LLM-based recommenders, leading not only to decreased recommendation accuracy but also to pronounced unfairness issues. On the one hand, critical user–item interaction signals that capture user preferences may be ignored by the model, severely impairing its ability to deliver personalized recommendations. On the other hand, the recommendation output becomes inherently skewed toward a limited subset of items whose textual tokens exhibit strong correlations with auxiliary tokens.

To empirically verify this phenomenon, we conduct experiments on the typical Amazon datasets. Specifically, we partition items into five groups based on their semantic relevance to auxiliary tokens, measured by the model’s estimated probability:
\begin{equation}
\begin{aligned}
\label{eq:item_relevance}
r(y) &= \frac{1}{|y|} \sum_{t=1}^{|y|} \log P_\theta(y_t | x^{\text{task}},y_{<t})
\end{aligned}
\end{equation}
Here we mask the user historical information in the original prompt template with the placeholder `N/A', and compute the probability of the tokens for each item. This measure serves as an indicator of the relevance between the input and output, a practice commonly adopted for estimating semantic relevance \cite{kauf2024log}.
Based on this metric, items are sorted and evenly divided into five groups, with Group 1 containing the items of highest relevance and Group 5 the least.

Next, we compute the proportion of Top-1 recommended items from each group, with the results shown in Figure \ref{fig:fairness_intro}. The analysis reveals that fine-tuned LLMs display a pronounced inclination for items in Group 1, whose tokens exhibit the highest relevance to auxiliary tokens. Nearly 80\% of Top-1 recommendations fall within this group, despite such items comprising only about 20\% of the test set. As illustrated in Figure \ref{fig:fairness_curve_intro}, this bias becomes progressively more pronounced as the tuning proceeds.
This bias not only undermines recommendation accuracy but also exacerbates exposure unfairness by systematically over-promoting items misaligned with individual user preferences. Such skew can severely degrade user experience and distort the recommendation ecosystem. For example, it would incentivize content providers to adopt clickbait-like titles or other superficial textual strategies to artificially increase token relevance to the auxiliary tokens.

\begin{figure}[t]
  \centering
  \includegraphics[width=0.85\linewidth]{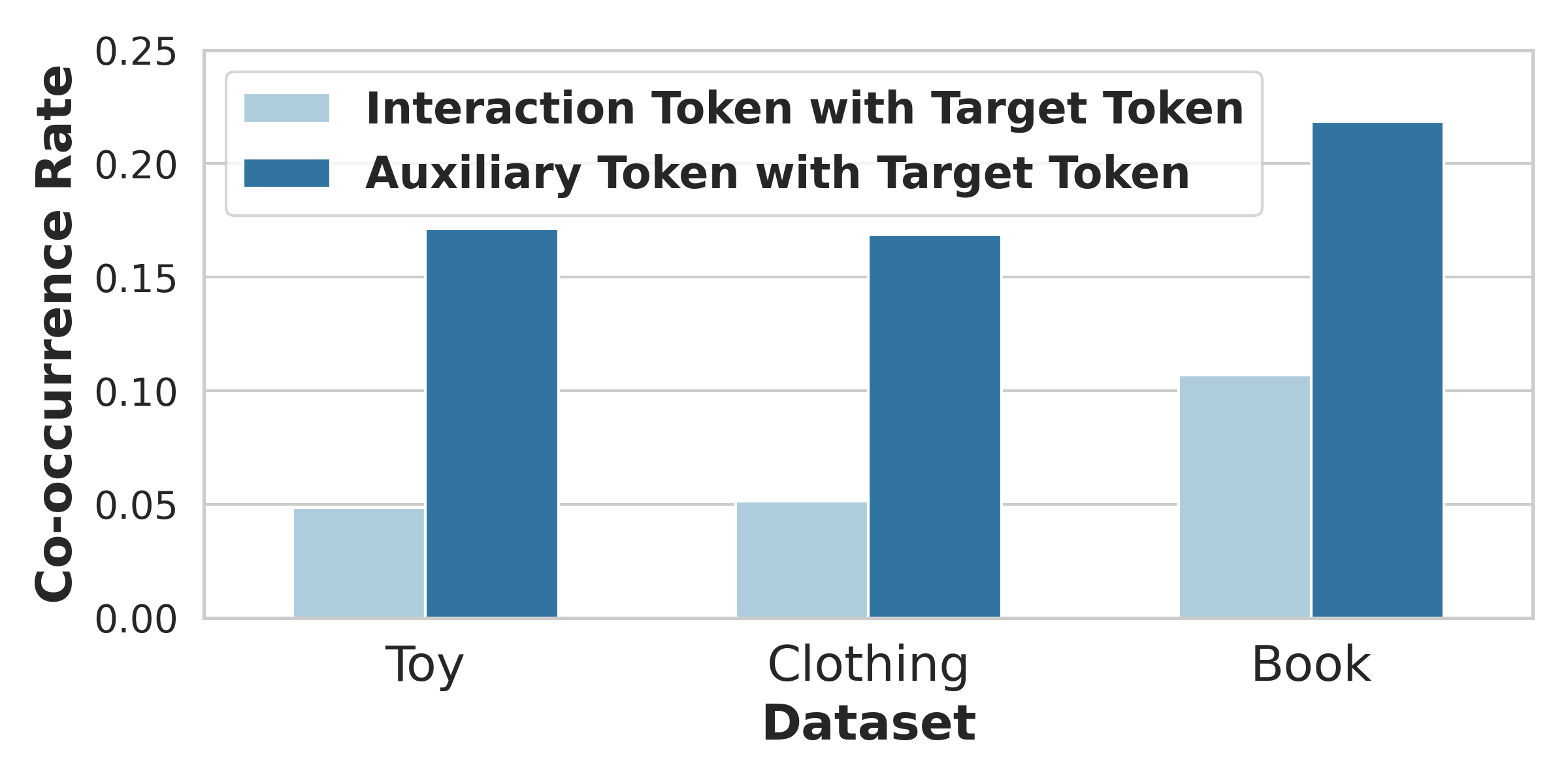}
  \caption{The co-occurrence rate of different types of token pairs in the training set.}
  \Description{}
  \label{fig:token_co_occurrence_rate}
\end{figure}

\subsubsection{Origins of Context Bias}
\label{sec:origins}
The emergence of context bias can be traced to biases inherent in the training data. As shown in Figure~\ref{fig:token_co_occurrence_rate}, the co-occurrence rate between auxiliary tokens and target item tokens is significantly higher than those between interaction tokens and target item tokens. This is expected, as task prompts appear in every training instance, and prefix tokens are always accompanied for item tokens. Consequently, during fine-tuning, the LLM tends to capture shortcut patterns, simply memorizing frequent correlations with auxiliary tokens, while neglecting the more important user-specific interaction signals. This incurs context bias and motivates the need for improved fine-tuning strategies.


\subsubsection{Limitations of Related Strategies}
\label{sec:limitations}
Several recent studies have explored bias and fairness issues in LLM-based RS. However, the characteristics of these biases differ from those of \textit{context bias}. Specifically, context bias arises during fine‑tuning and reflects the inherent over‑reliance of LLMs on auxiliary tokens.
In contrast, earlier work has primarily focused on biases such as \textit{popularity bias} \cite{gao2025process, lu2025dual, jiang2024item, gao2025sprec, liao2024rosepo}, arising from imbalanced item frequency in training data; \textit{position bias} \cite{chen2023bias, ma2023large, hou2024large, bito2025evaluating, jiang2025beyond, dai2024bias, luo2025recranker, zhang2024agentcf}, caused by the model’s sensitivity to the ordering of candidate items; \textit{amplification bias} \cite{bao2024decoding}, arising from length normalization, which favors items containing tokens with generation probabilities close to 1.
These biases stem from distinct sources and mechanisms.  
Consequently, this newly identified form of context bias warrants explicit and targeted mitigation, yet recent work has fallen short in addressing this issue (\cf~Table \ref{tab:cmp_exp}).

Another related work, CFT \cite{zhang2024causality}, aims to encourage LLMs to better leverage user interaction information.  
CFT introduces counterfactual learning to forcibly enhance the influence of interaction tokens for each training instance. However, this strategy suffers from multiple aspects of limitations: \textbf{(1) Objective misalignment}. The counterfactual objective is not directly aligned with improving recommendation accuracy or fairness, and its contribution to these aspects remains uncertain. In fact, CFT only treats this objective as an auxiliary loss and over-emphasis on this term has been observed to lead to substantial performance degradation. \textbf{(2) Difficulty in weight selection}. CFT relies on manually specified weights to determine the degree to which a training instance should rely on interaction tokens. In fact, the optimal weights can vary substantially across instances and are difficult to estimate accurately. Manual specification is therefore challenging, prone to deviation from the ideal value, and ultimately detrimental to model effectiveness. Empirically, even when we directly employ the official source code of CFT and conduct a fine‑grained hyperparameter search, CFT yields only limited performance gains on some datasets (\cf~Table \ref{tab:cmp_exp}). \textbf{(3) High computational overhead}. CFT requires to process counterfactual instances, resulting in more than double the training time compared with SFT (\cf Figure~\ref{fig:efficiency}).
In contrast, our GDRT is explicitly aligned with the target objective, enabling the model to consistently perform well across different instance groups. This naturally encourages greater focus on interaction tokens without requiring manual specification of influence strengths.

\section{Methodology}
The above analyses reveal a significant context bias inherent in fine-tuning LLMs for recommendation. To address this issue, we propose a novel fine‑tuning framework, termed GDRT, which leverages Group Distributionally Robust Optimization (Group DRO) \cite{sagawa2019distributionally} to mitigate such bias. This section first outlines the general idea of GDRT, and then describes the group partitioning strategy and customized loss function.

\textbf{General Idea.} Rather than directly intervening the learning process to manually enhance the effect of interaction tokens, we pursue an alternative objective: \emph{ensuring that the model achieves consistently strong performance across target tokens regardless of their degree of relevance to auxiliary tokens.} It naturally shifts the LLM's attention away from auxiliary tokens towards user‑specific interaction tokens, as simple reliance on shortcut auxiliary tokens results in poor performance on the instances with weak correlations. Besides, this objective is directly aligned with the goals of high recommendation accuracy and fairness. 

To implement this idea, we adopt Group DRO, which partitions the training data into multiple groups and uses an adversarial training mechanism to encourage great performance across all groups. Group DRO has been widely applied in various domains and shown effectiveness in mitigating group disparities and shortcut correlations \cite{sagawa2019distributionally, zhao2023popularity, oren2019distributionally}. In applying Group DRO to our problem, we address two key questions: (1) how to construct groups that capture varying degrees of relevance to auxiliary tokens; and (2) how to design a loss function that is computationally efficient.

\textbf{Token Grouping by Relevance with Auxiliary Tokens.}
Our objective is to ensure consistent model performance across groups that differ in their degree of relevance to auxiliary tokens. Accordingly, the constructed groupings should capture variations in this relevance.
Towards this end, we compute the predictive probability of each target token conditioned solely on the corresponding auxiliary tokens:  
\begin{equation}
\label{eq:token_relevance}
r(y^*_{i,t}) = \log P_\theta\left(y^*_{i,t} \mid x^{\text{task}}, y^*_{i,<t}\right).
\end{equation}
where \(y^*_{i,t}\) denotes the \(t\)-th token of the target item in the \(i\)-th training sample, and \(y^*_{i,<t}\) denotes its prefix tokens.
Based on these relevance scores, all target tokens are partitioned into $G$ disjoint groups $\{\mathcal{G}_1, \dots, \mathcal{G}_G\}$ using the K-means algorithm\footnote{We can simply employ the \textit{scikit-learn} package to implement the K-means algorithm. Considering that \( r(y^*_{i,t}) \) is a numerical value, it can be evenly partitioned according to its magnitude, yielding comparable results.}, where $G$ is a hyperparameter that determines the granularity of grouping. Each group $\mathcal{G}_g$ contains the index $(i,t)$ of target tokens exhibiting similar degrees of relevance to auxiliary tokens, whereas tokens belonging to different groups demonstrate distinct relevance levels.

Unlike Eq.~\ref{eq:item_relevance}, which evaluates \emph{item-wise} relevance between the entire target item and the auxiliary tokens, Eq.~\ref{eq:token_relevance} computes \emph{token-wise} relevance. This distinction is crucial, as different tokens within the same target item may exhibit varying degrees of relevance to the auxiliary tokens. Such finer-grained grouping can more effectively capture variations in this relevance.

\begin{figure}[t]
  \centering
  \includegraphics[width=0.8\linewidth]{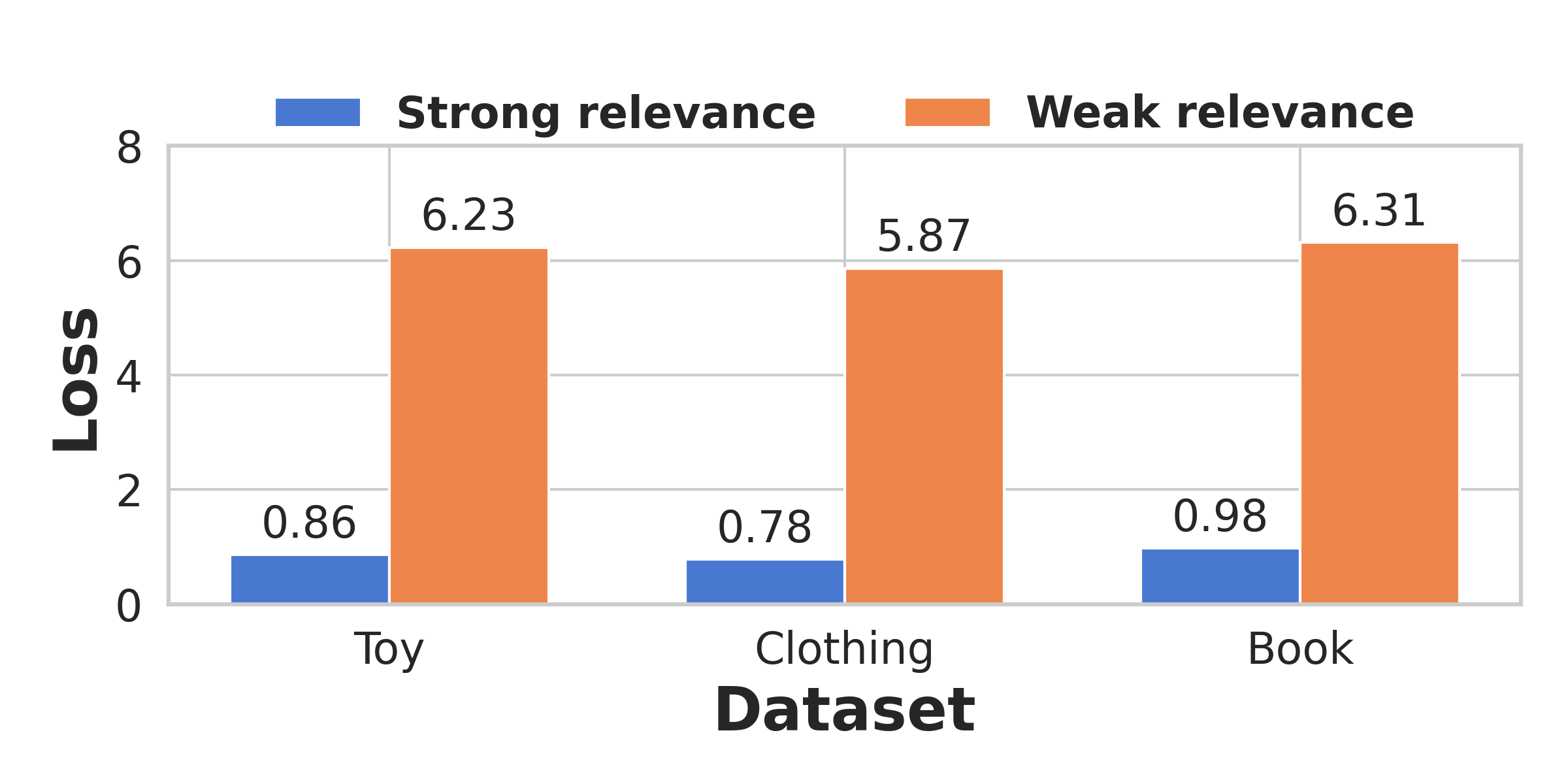}
  \caption{Loss of token groups with varying degrees of relevance to auxiliary tokens in the SFT-trained biased model. Target tokens are grouped into strong- and weak-relevance sets using K-means on relevance scores computed with Eq.~\ref{eq:token_relevance}.}
  \Description{}
  \label{fig:group_correlation_loss}
\end{figure}

\textbf{The Objective of GDRT.} 
Subsequently, we employ Group DRO to enforce consistent model performance across groups with different strength of relevance. Specifically, the training objective of GDRT is defined as:
\begin{equation}
\begin{aligned}
\label{eq:GDRT_loss}
&\mathcal{L}_{GDRT} = \max_{Q} \sum_{g=1}^{G} Q(g) \mathcal{L}(g) \quad
\text{s.t.} \quad D_{KL}({Q},{U})\leq\eta \\
&\mathcal{L}(g) = \frac{1}{|\mathcal{G}_g|}\sum_{(i,t) \in \mathcal{G}_g} -\log P_\theta \left(y^*_{i,t} \mid x_i, y^*_{i,<t}\right)
\end{aligned}
\end{equation}
where $\mathcal{L}(g)$ represents the vanilla generative loss for group $\mathcal{G}_g$, and \(|\mathcal{G}_g|\) represents the number of target tokens in group \(\mathcal{G}_g\). The term $Q$ denotes the weight distribution over groups, acting as a flexible adversarial perturbation to the original empirical distribution, with $Q(g)$ being the weight assigned to the $g$-th group. This weight distribution is regularized via a Kullback–Leibler divergence term \(D_{KL}(Q, U)\) between the weight distribution $Q$ and the uniform distribution $U$, with the parameter $\eta$ controlling the perturbation magnitude. 

Comparing $\mathcal{L}_{GDRT}$ (Eq.~\ref{eq:GDRT_loss}) with the original SFT objective \( \mathcal{L}_{SFT} \) (Eq.~\ref{eq:SFT_loss}), the main difference lies in the introduction of the additional group weighting term $Q(g)$. Intuitively, Group DRO imposes an adversarial shift to the group distribution by perturbing the relevance distribution with respect to auxiliary tokens, thereby compelling the model to perform consistently well across groups under different adversarial re‑weightings. This mechanism naturally mitigates the model's reliance on auxiliary tokens, and encourages the model to perform consistently well across different groups. 

\textbf{Efficient Implementation.} 
While Group DRO involves the complex adversarial optimization,  it can be simplified to an equivalent closed‑form objective. We have the following lemma:
\begin{lemma}
\label{lemma1}
    Equation \ref{eq:GDRT_loss} can be reformulated as the following objective:
    \begin{equation}
    \begin{aligned}
    \label{eq:eff_gdrt_loss}
    \mathcal{L}_{GDRT} = \sum_{g=1}^{G} Q(g) \mathcal{L}(g), \quad Q(g) = \frac{e^{\mathcal{L}(g)/\tau}}{\sum_{g'=1}^{G} e^{\mathcal{L}(g')/\tau}}
    \end{aligned}
    \end{equation}
    The parameter \( \tau \) is the dual Lagrange coefficient associated with the constraint \( D_{KL}(Q, U)\leq\eta \).
\end{lemma}

The lemma yields an explicit closed‑form solution for the group weights $Q(g)$. From the perspective of these weights, the effect of Group DRO can be well understood: groups with higher losses, which typically correspond to tokens exhibit weaker relevance with auxiliary tokens (see Figure~\ref{fig:group_correlation_loss}), receive larger weights. This adaptive reweighting encourages the model to allocate greater optimization emphasis to underperforming groups, thereby enhancing its learning on samples with low auxiliary-token relevance. As a result, the model attains more balanced performance across groups and effectively mitigates its over-reliance on auxiliary tokens.

This closed‑form also facilitates an efficient implementation of GDRT. Compared to SFT, the additional steps just involve: partitioning the token groups based on Eq.~\ref{eq:token_relevance}, and dynamically updating the group weights according to Eq.~\ref{eq:eff_gdrt_loss}. The time complexity of GDRT is the same as the basic SFT, and our empirical experiments also demonstrate their close running time (\cf~Figure \ref{fig:efficiency}).
Besides, this simple reformulation makes the integration of GDRT into existing LLM-based RS straightforward, requiring only minimal code changes. 

\begin{table}[t]
\centering
\caption{Statistics of the datasets.}
\label{tab:statistics}
{%
\begin{tabular}{@{}lccccc@{}}
\toprule
Dataset  & \#Users & \#Items & \#Interactions & \#Density  \\ \midrule
Toy      & 19124  & 11758  & 165247        & 0.0735\% \\
Clothing & 39230  & 22948  & 277534        & 0.0308\% \\
Book     & 16559  & 6344   & 151928        & 0.1446\% \\ \bottomrule
\end{tabular}%
}
\end{table}

\begin{table*}[t]
\centering
\caption{The performance comparison on three real-world datasets. The best result is bolded. Lower MGU and DGU indicate better fairness.}
\label{tab:cmp_exp}
\begin{tabular}{@{}l|l|cccccccc
>{\columncolor[HTML]{C0F1FF}}c @{}}
\toprule
Dataset                    & Metric             & SASRec & DROS   & SASRec++ & SFT    & Reweight & D3     & SPRec  & CFT    & GDRT            \\ \midrule
                           & NDCG@5 $\uparrow$  & 0.0057 & 0.0095 & 0.0120   & 0.0118 & 0.0084   & 0.0115 & 0.0148 & 0.0119 & \textbf{0.0152} \\
                           & NDCG@10 $\uparrow$ & 0.0073 & 0.0118 & 0.0144   & 0.0158 & 0.0121   & 0.0160 & 0.0175 & 0.0158 & \textbf{0.0203} \\
                           & HIT@5 $\uparrow$   & 0.0090 & 0.0167 & 0.0200   & 0.0202 & 0.0146   & 0.0186 & 0.0234 & 0.0188 & \textbf{0.0246} \\
                           & HIT@10 $\uparrow$  & 0.0138 & 0.0240 & 0.0271   & 0.0330 & 0.0261   & 0.0325 & 0.0317 & 0.0311 & \textbf{0.0405} \\
                           & MGU@5 $\downarrow$ & /      & /      & /        & 0.1870 & 0.2119   & 0.1740 & 0.2255 & 0.1964 & \textbf{0.1288} \\
\multirow{-6}{*}{Toy}      & DGU@5 $\downarrow$ & /      & /      & /        & 0.6122 & 0.6655   & 0.5722 & 0.7197 & 0.6257 & \textbf{0.4031} \\ \midrule
                           & NDCG@5 $\uparrow$  & 0.0016 & 0.0043 & 0.0024   & 0.0038 & 0.0038   & 0.0049 & 0.0045 & 0.0042& \textbf{0.0054} \\
                           & NDCG@10 $\uparrow$ & 0.0024 & 0.0053 & 0.0033   & 0.0063 & 0.0066   & 0.0075 & 0.0072 & 0.0068& \textbf{0.0096} \\
                           & HIT@5 $\uparrow$   & 0.0032 & 0.0080 & 0.0042   & 0.0078 & 0.0074   & 0.0100 & 0.0084 & 0.0086& \textbf{0.0118} \\
                           & HIT@10 $\uparrow$  & 0.0058 & 0.0110 & 0.0070   & 0.0156 & 0.0160   & 0.0180 & 0.0168 & 0.0168& \textbf{0.0246} \\
                           & MGU@5 $\downarrow$ & /      & /      & /        & 0.1553 & 0.2491   & 0.1211 & 0.1728 & 0.1410& \textbf{0.0522} \\
\multirow{-6}{*}{Clothing} & DGU@5 $\downarrow$ & /      & /      & /        & 0.4908 & 0.7988   & 0.4069 & 0.5443 & 0.4686& \textbf{0.1996} \\ \midrule
                           & NDCG@5 $\uparrow$  & 0.0054 & 0.0060 & 0.0065   & 0.0067 & 0.0051   & 0.0073 & 0.0033 & 0.0080 & \textbf{0.0139} \\
                           & NDCG@10 $\uparrow$ & 0.0071 & 0.0084 & 0.0081   & 0.0103 & 0.0070   & 0.0112 & 0.0071 & 0.0105 & \textbf{0.0145} \\
                           & HIT@5 $\uparrow$   & 0.0089 & 0.0110 & 0.0100   & 0.0116 & 0.0100   & 0.0130 & 0.0057 & 0.0137 & \textbf{0.0226} \\
                           & HIT@10 $\uparrow$  & 0.0141 & 0.0185 & 0.0153   & 0.0228 & 0.0160   & 0.0240 & 0.0178 & 0.0219 & \textbf{0.0244} \\
                           & MGU@5 $\downarrow$ & /      & /      & /        & 0.1433 & 0.1382   & 0.1229 & 0.1709 & 0.1657 & \textbf{0.0579} \\
\multirow{-6}{*}{Book}     & DGU@5 $\downarrow$ & /      & /      & /        & 0.4742 & 0.3406   & 0.4238 & 0.5477 & 0.4643 & \textbf{0.2240} \\ \bottomrule
\end{tabular}
\end{table*}

\begin{table}[t]
\centering
\caption{Performance comparison under different LLM-based RS. The best result is bolded.}
\label{tab:gen_exp}
\begin{tabular}{@{}l|cc|cc@{}}
\toprule
\multirow{2}{*}{Method} & \multicolumn{2}{c|}{Toy}          & \multicolumn{2}{c}{Clothing}      \\ \cmidrule(l){2-5} 
                        & NDCG@5          & DGU@5           & NDCG@5          & DGU@5           \\ \midrule
MSL                     & 0.0198          & 0.7162          & 0.0077          & 0.5815          \\
MSL+GDRT                & \textbf{0.0253} & \textbf{0.2697} & \textbf{0.0096} & \textbf{0.1469} \\ \midrule
LLaRA                   & 0.0131          & 0.6485          & 0.0041          & 0.5412          \\
LLaRA+GDRT              & \textbf{0.0168} & \textbf{0.4249} & \textbf{0.0053} & \textbf{0.2320} \\ \midrule
A-LLM                   & 0.0129          & 0.6122          & 0.0045          & 0.5238          \\
A-LLM+GDRT              & \textbf{0.0160} & \textbf{0.4192} & \textbf{0.0053} & \textbf{0.1817} \\ \bottomrule
\end{tabular}
\end{table}

\section{Experiments}
We aim to answer the following research questions:
\begin{itemize}[left=5pt]
\item {RQ1: How does GDRT perform compared to SOTA methods?} 
\item {RQ2: Can GDRT be integrated into other LLM-based RS?} 
\item {RQ3: Does GDRT mitigate context bias?} 
\item {RQ4: How do the hyperparameters affect the GDRT?} 
\item {RQ5: How does the efficiency of GDRT compare with baselines?} 
\end{itemize}

\subsection{Experimental Settings}
\subsubsection{Datasets}
Three widely used real-world datasets—\textit{Amazon Toys and Games}, \textit{Amazon Clothing, Shoes and Jewelry}, and \textit{Amazon Books}\footnote{\url{https://jmcauley.ucsd.edu/data/amazon/index_2014.html}}—are employed in our experiments. 
To ensure fair comparison, we follow the data preprocessing procedures employed in recent literature \cite{bao2025bi, cui2024distillation, wang2025msl}. 
Specifically, we first apply the 5-core setting to the raw datasets. For user interaction sequences exceeding 11 interactions, we segment the sequences using a sliding window of length 11. The segmented sequences are then sorted in ascending order by timestamp and partitioned into training, validation, and test sets in an 8:1:1 ratio. Due to the large size of \textit{Amazon Books}, we randomly retain 100,000 items prior to 5-core processing. 
The statistics of the processed datasets are summarized in Table \ref{tab:statistics}.

\subsubsection{Baselines}
The methods compared fall into several categories: 
\textbf{(1) Traditional RS}: \underline{SASRec} \cite{kang2018self}, \underline{SASRec++} \cite{lepage2025closing}, \underline{DROS} \cite{yang2023generic}. 
\textbf{(2) LLM-based RS}: \underline{SFT} \cite{bao2025bi}, \underline{CFT} \cite{zhang2024causality}, \underline{MSL} \cite{wang2025msl}, \underline{LLaRA} \cite{liao2024llara}, \underline{A-LLM} \cite{kim2024large}. 
\textbf{(3) Debiasing for LLM-based RS}: \underline{Reweight} \cite{jiang2024item}, \underline{SPRec} \cite{gao2025sprec}, \underline{D3} \cite{bao2024decoding}. For a detailed description, see Appendix \ref{apd:baseline}.


\subsubsection{Implementation Details} 
\label{sec:implementation_details}
For all LLM-based methods, we adopt LLaMA3.2-3B \cite{dubey2024llama} as the backbone, with the number of training epochs set to 5. The prompt design follows \cite{bao2025bi}.
For inference, we consistently employ Constrained Beam Search (CBS) across all baselines, following prior work \cite{wang2025msl, bao2024decoding}, to ensure that the recommended items are drawn from the item set. The beam size is fixed at 10.  
For evaluation, we evaluate the NDCG@5 on the validation set for each epoch's checkpoint and select the checkpoint with the highest score for testing. The corresponding results on the test set are reported as the final performance. 
For the hyperparameters in GDRT, the number of groups \( G \) is tuned from \(\{2, 5, 10\}\), and \(\tau\) is tuned from \(\{0.1, 0.2, 0.3, 0.5, 1.0\}\).
To ensure fair comparisons, we utilize the source code provided by the original authors and tune the hyperparameters of all baseline methods according to the guidelines specified in their respective publications.

\subsubsection{Metrics}
Four widely used evaluation metrics are employed in this study: \textit{NDCG@K} and \textit{Hit Ratio@K} are used to evaluate recommendation accuracy, while \textit{MGU@K} and \textit{DGU@K} are adopted to evaluate fairness \cite{jiang2024item} (K=5, 10). Specifically, for the fairness metrics, items are divided into five groups according to Equation~\ref{eq:item_relevance}. We then calculate the discrepancy between each group’s proportion in the Top-K recommendations and its proportion in the user's interaction history. \textit{MGU@K} measures the average of these discrepancies, whereas \textit{DGU@K} quantifies the gap between the maximum and minimum discrepancies across groups. Accordingly, smaller values of \textit{MGU@K} and \textit{DGU@K} indicate that the distribution of recommended items is more aligned with that of the user’s historical interactions, reflecting better fairness.

\subsection{Performance Comparison (RQ1 \& RQ2)}
Table \ref{tab:cmp_exp} presents a comparative analysis of the proposed GDRT method against the baselines. Overall, GDRT demonstrates substantial performance improvements across all datasets.  
By mitigating context bias, reducing excessive reliance on auxiliary tokens, and enhancing the model’s ability to capture user-specific behavioral patterns, GDRT delivers marked gains in both accuracy and fairness. 
Specifically, GDRT achieves an average improvement of 24.29\% in NDCG@10 and 37.43\% reduction in MGU@5 compared with the best baseline. 
In contrast, other comparison methods yield only marginal improvements over SFT, and their lack of explicit mechanisms to address context bias results in limited fairness enhancement.
Furthermore, since GDRT operates solely by modifying the loss function, it can be easily integrated into existing LLM-based RS without altering their architectures. To verify its general applicability, we further evaluate GDRT when incorporated into several advanced LLM-based recommendation methods. As shown in Table \ref{tab:gen_exp}, GDRT consistently yields significant improvements in both performance and fairness across all evaluated methods, demonstrating its strong generalization capability.
Additional comparisons using alternative prompt templates and LLMs beyond those used in the main experiments are provided in Appendix \ref{apd:performance}.

\begin{figure}[t]
  \centering
  \includegraphics[width=\linewidth]{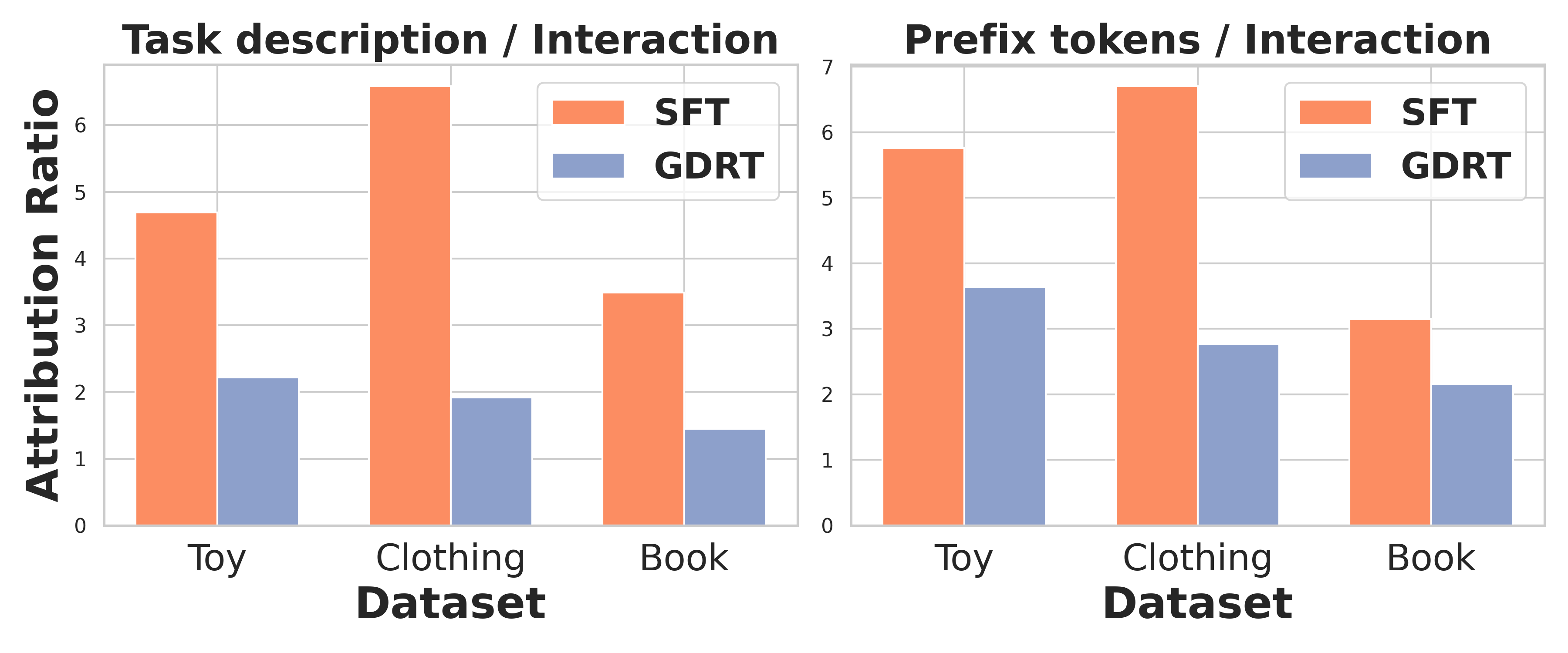}
  \caption{Ratio of attribution values between auxiliary tokens and user interaction tokens using SFT and GDRT. Left: task description vs. user interaction tokens. Right: prefix tokens of predicted item (take the first token) vs. user interaction tokens.}
  \Description{}
  \label{fig:feature_ablation_exp}
\end{figure}



\begin{figure}[t]
  \centering
  \includegraphics[width=\linewidth]{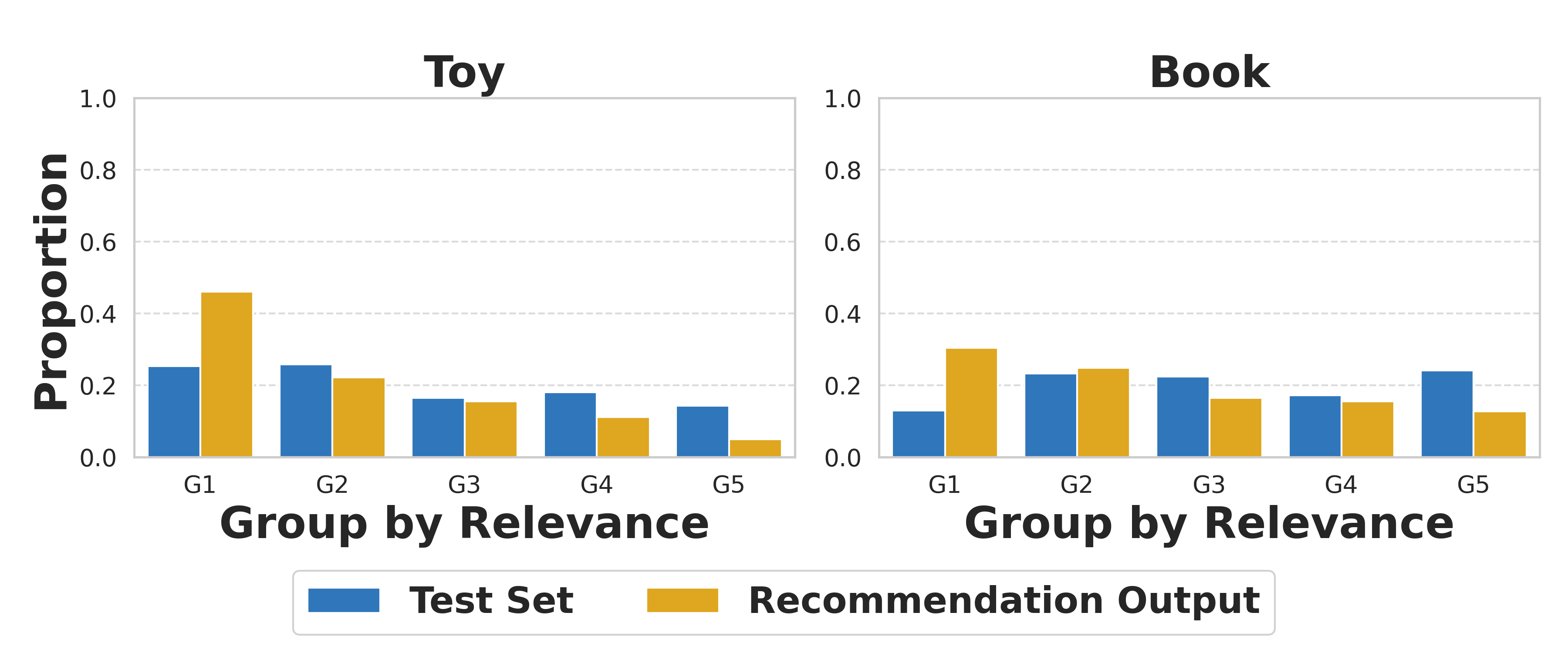}
  \caption{Distribution of Top‑1 recommended items from the GDRT-trained model across five groups defined by auxiliary-token relevance (Group 1: highest relevance, Group 5: lowest relevance), with test-set distribution for comparison.}
  \Description{}
  \label{fig:fairness_exp}
\end{figure}

\subsection{In-depth Analysis (RQ3)}
\begin{figure}[t]
  \centering
  \includegraphics[width=\linewidth]{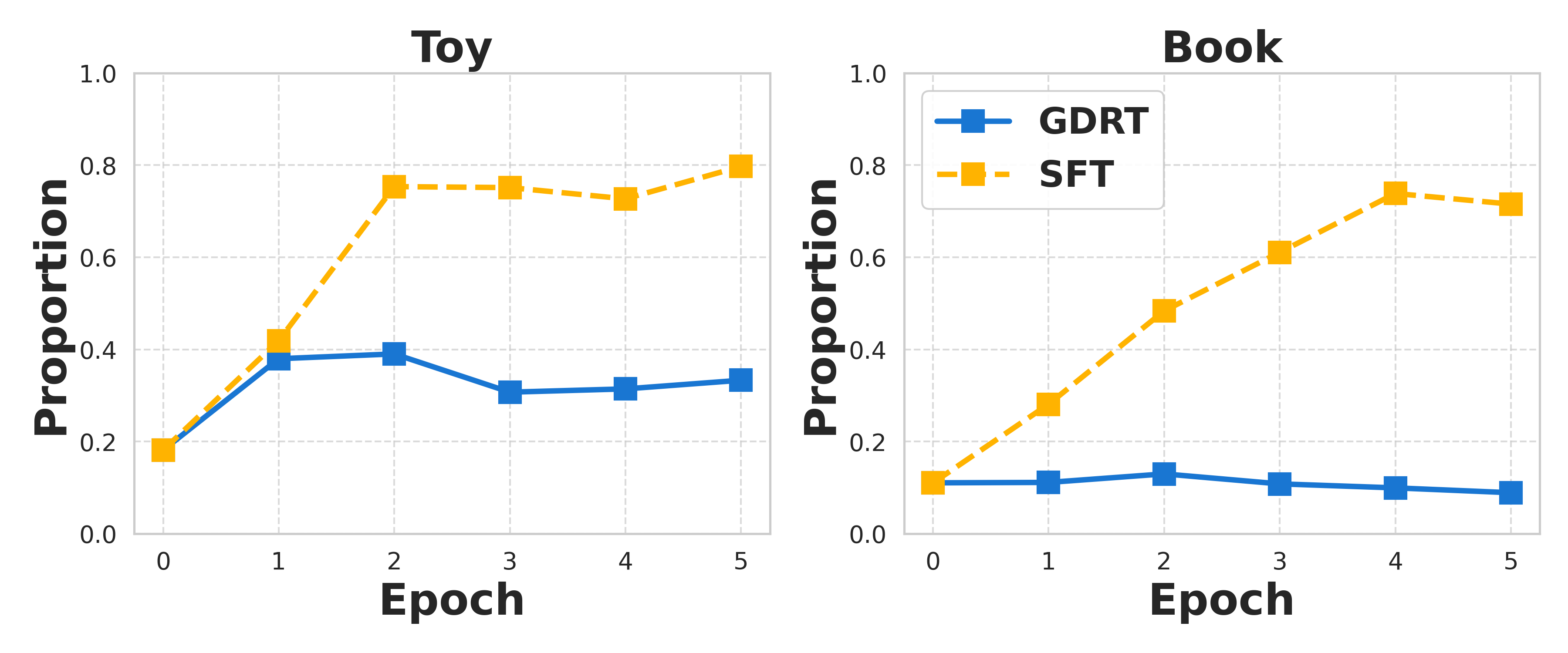}
  \caption{Proportion of Top‑1 recommendations belonging to Item Group 1 (highest relevance with auxiliary tokens) during training with SFT and GDRT.}
  \Description{}
  \label{fig:fairness_curve_exp}
\end{figure}

In this section, we present an empirical analysis of the effectiveness of GDRT in mitigating context bias. 
First, we perform FAA on both SFT- and GDRT-trained models, measuring the ratio of attribution values assigned to auxiliary tokens versus interaction tokens. As shown in Figure~\ref{fig:feature_ablation_exp}, GDRT yields a significantly lower ratio than SFT, indicating that GDRT effectively alleviates the model’s over-reliance on auxiliary tokens.
Furthermore, we analyze the recommendation distributions on different item groups with varying degrees of auxiliary-token relevance. As illustrated in Figure~\ref{fig:fairness_exp}, the recommendation distribution generated by the GDRT-trained model closely align with the distribution of the test set. This improvement can be attributed to GDRT’s ability to suppress the tendency of SFT to progressively amplify items highly correlated with auxiliary tokens as shown in Figure~\ref{fig:fairness_curve_exp}.
Overall, these results demonstrate that GDRT effectively mitigates context bias.


\begin{figure}[t]
  \centering
  \includegraphics[width=\linewidth]{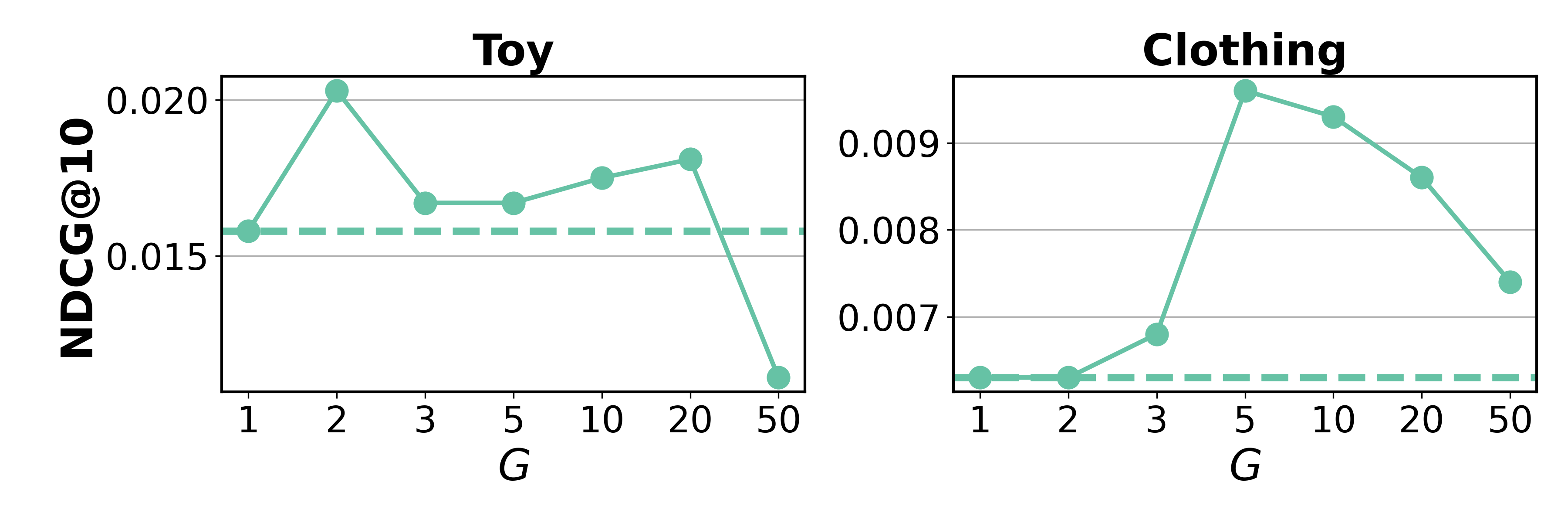}
  \caption{Hyperparameter sensitivity analysis on group number $G$ (dashed: SFT baseline).}
  \Description{}
  \label{fig:hyperparameter_sensitivity_groupnum}
\end{figure}

\begin{figure}[t]
  \centering
  \includegraphics[width=\linewidth]{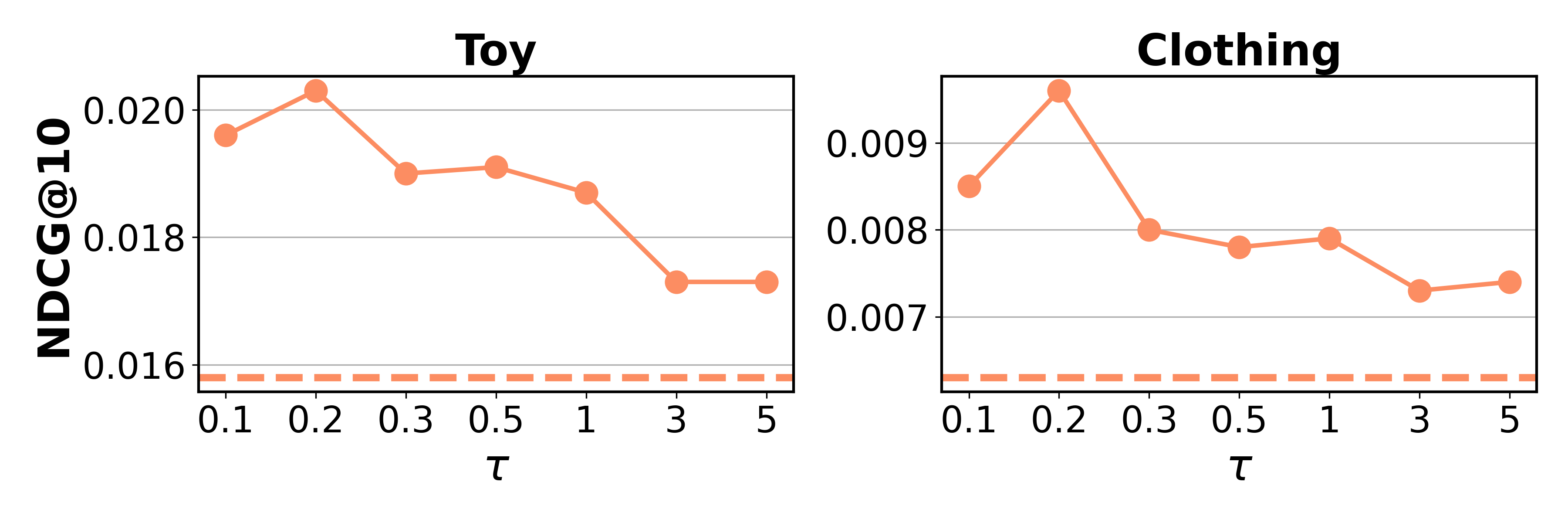}
  \caption{Hyperparameter sensitivity analysis on parameter $\tau$ in DRO (dashed: SFT baseline).}
  \Description{}
  \label{fig:hyperparameter_sensitivity_tau}
\end{figure}

\subsection{Hyper-Parameter Sensitivities (RQ4)}
In this section, we perform a sensitivity analysis on the hyperparameters of GDRT: the number of groups \(G\) in K-means (Figure \ref{fig:hyperparameter_sensitivity_groupnum}) and the coefficient \(\tau\) in DRO (Figure \ref{fig:hyperparameter_sensitivity_tau}).
For both parameters, model performance exhibits a trend of initially increasing and then decreasing as the parameter values rise. This phenomenon can be explained as follows. 
For \(G\), an excessively small value may group highly heterogeneous data together, thereby diminishing the differences in auxiliary-token relevance between groups. Conversely, an overly large \(G\) results in groups containing too few samples to reliably represent the underlying distribution.
As for \(\tau\), a smaller value places greater emphasis on groups with higher losses; however, overemphasizing the worst-performing group can cause overfitting and impair overall generalization. In contrast, a larger \(\tau\) balances optimization across all groups, but may reduce group-wise consistency in performance.
Besides, the model demonstrates strong performance across a wide range of parameter settings, indicating the robustness of GDRT to hyperparameter selection.

\begin{figure}[t]
  \centering
  \includegraphics[width=\linewidth]{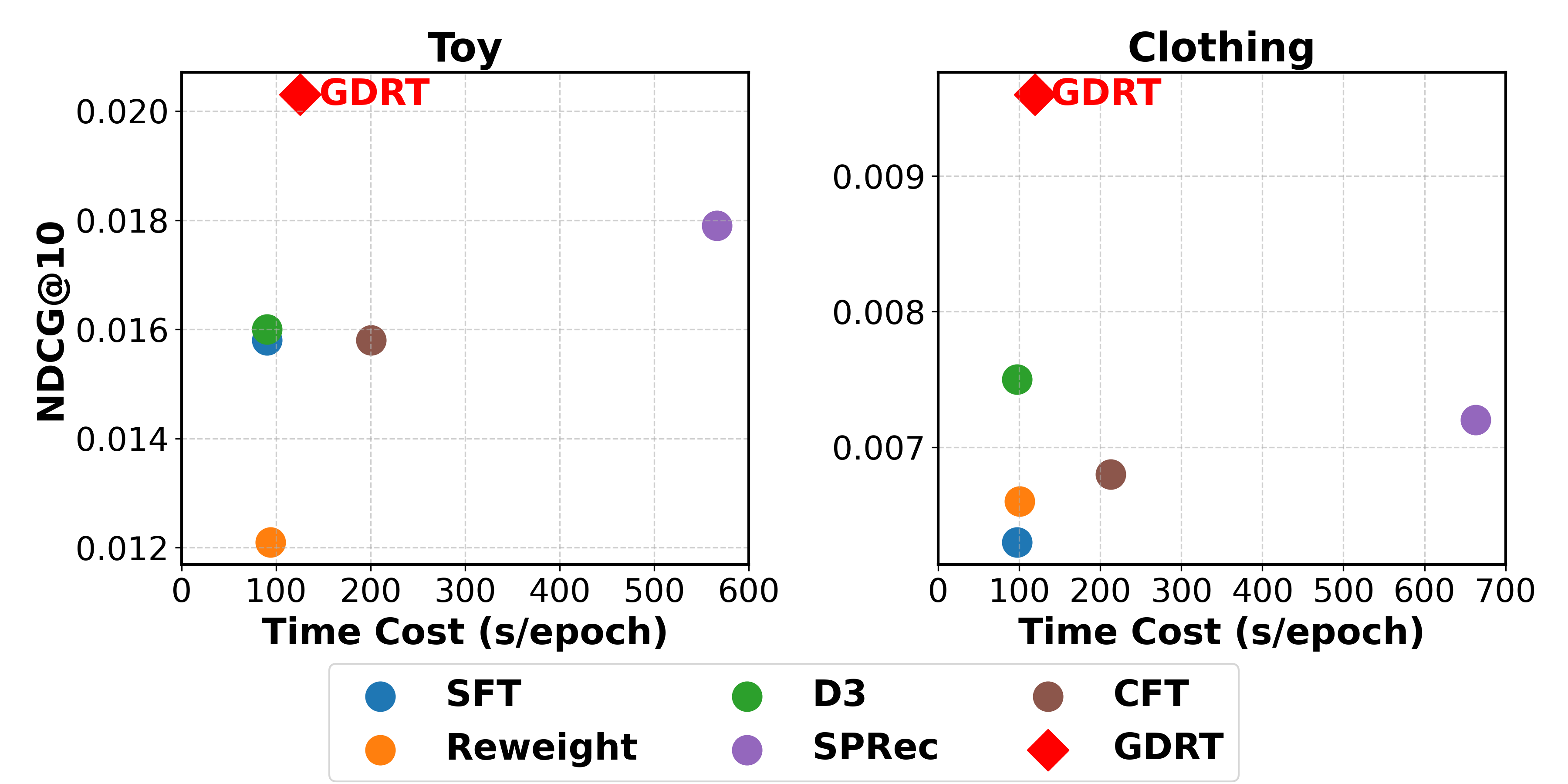}
  \caption{Performance comparisons in terms of both recommendation accuracy and efficiency.}
  \Description{}
  \label{fig:efficiency}
\end{figure}

\subsection{Efficiency Comparison (RQ5)}
\label{sec:efficiency}
This section compares the efficiency and performance of GDRT with baselines. As illustrated in Figure \ref{fig:efficiency}, GDRT demonstrates both optimal performance and high computational efficiency, incurring negligible overhead compared to SFT.
In contrast, CFT introduces additional counterfactual samples leading to substantially higher computational costs. SPRec, based on DPO \cite{rafailov2024direct}, requires an extra reference model and negative samples during training, resulting in a considerable increase in runtime. Although Reweight and D3 are relatively efficient, their performance improvements are limited.


\section{Related Work}
\subsection{Sequential Recommendation}
Sequential recommendation focuses on predicting the next item a user will be interested in based on their historical interactions. Compared to collaborative filtering \cite{chen2025rankformer, chen2024sigformer, yang2024psl, zhang2025advancing, yang2025breaking, wang2024distributionally}, sequential recommendation incorporates temporal information and places greater emphasis on capturing the evolving patterns of user interests. 
With the advancement of deep learning, numerous architectures based on deep neural networks have been introduced into the sequential recommendation. For example, GRU4Rec \cite{hidasi2015session} employs RNNs, while Caser \cite{tang2018personalized} utilizes CNNs to effectively capture long-term dependencies and modeling user interest patterns from historical behavior. 
More advanced models such as SASRec \cite{kang2018self} and BERT4Rec \cite{sun2019bert4rec} leverage self-attention mechanisms \cite{vaswani2017attention}, enabling the identification of the most relevant parts within the sequence. Due to the dynamic nature of data distributions as time evolves \cite{wang2024cola, wang2024spatiotemporal}, DROS \cite{yang2023generic} introduces DRO \cite{rahimian2019distributionally, wu2023understanding} to further enhance the model's robustness against distributional shifts caused by temporal changes. The readers may refer to the survey \cite{fang2020deep} for more details.

\subsection{Biases in LLM-based Recommendation}
Large Language Models (LLMs), with their powerful capabilities in comprehension, reasoning, and extensive knowledge \cite{dubey2024llama, achiam2023gpt, guo2025deepseek, team2024qwen2}, have been widely applied to recommendation systems \cite{wu2024survey, wang2025llm4dsr, cui2025field, cui2025hatllm}. One prominent paradigm is LLM-based RS \cite{li2023large}, which directly leverages LLMs as the backbone of the recommender. Subsequent studies have explored fine-tuning LLMs on domain-specific recommendation datasets to further enhance their recommendation capabilities \cite{li2023large, bao2025bi, bao2023tallrec, zhu2024collaborative, wang2025msl, wang2024flip, liao2024llara, kim2024large, chen2024softmax}.

Recent studies have extensively explored bias and fairness issues in LLM-based RS, such as popularity bias \cite{gao2025process, lu2025dual, jiang2024item, gao2025sprec, liao2024rosepo, lin2025recommendation}, position bias \cite{chen2023bias, ma2023large, hou2024large, bito2025evaluating, jiang2025beyond, dai2024bias, luo2025recranker, zhang2024agentcf}, amplification bias \cite{bao2024decoding}, and bias stemming from LLMs’ preferences for specific item attributes \cite{li2023preliminary, deldjoo2024understanding, jiang2024item, zhang2025bifair}. 
Nevertheless, existing research has largely overlooked context bias, which arises during fine-tuning and reflects the inherent over-reliance of LLMs on auxiliary tokens. Since existing debiasing methods fail to account for this factor, their effectiveness remains limited. 
CFT \cite{zhang2024causality} seeks to enhance the modeling of users’ historical interactions, but suffers from objective misalignment, weight selection challenges, and high computational cost, restricting its applicability. These limitations are further discussed in Section~\ref{sec:limitations}.

\subsection{Group Distributionally Robust Optimization}
Group Distributionally Robust Optimization (Group DRO) \cite{sagawa2019distributionally} is an optimization framework that operates over predefined sample groups, aiming to achieve consistent and reliable performance across them by emphasizing the optimizing of the worst-performing group during training. It has been widely applied in various domains \cite{sagawa2019distributionally, zhao2023popularity, oren2019distributionally} and has demonstrated strong effectiveness in mitigating group disparities \cite{hashimoto2018fairness, zhao2023popularity} as well as reducing models’ reliance on shortcut correlations \cite{han2024mitigating, creager2021environment}.
Several studies have applied Group DRO to RS. For example, S-DRO \cite{wen2022distributionally} uses group DRO to improve the experience of underrepresented user groups that tend to engage with less popular items. PDRO \cite{zhao2023popularity} extends this approach with popularity-aware mechanisms to prevent harming the performance of popular items.



\section{Conclusion}
In this work, we identify a key limitation of supervised fine-tuning (SFT) in LLM-based recommenders: it often induces \textbf{Context Bias}, whereby the model over-relies on auxiliary tokens (\eg task descriptions and prefix-generated tokens) while underutilizing core user interaction information. This bias undermines recommendation accuracy and raises unfairness concerns.
To address this issue, we introduce {Group Distributionally Robust Optimization-based Tuning (GDRT)}, which aims to reduce the model’s over-reliance on auxiliary tokens by applying Group DRO across token groups with varying degrees of relevance to auxiliary tokens.
Extensive experiments on multiple public datasets demonstrate that GDRT effectively mitigates context bias, thereby significantly improving recommendation accuracy and enhancing fairness.

This work investigates a novel form of bias introduced by the integration of LLMs into recommenders, which is not present in traditional recommendation models. A promising avenue for future work is to examine whether LLM-based RS exhibit additional, as-yet unidentified biases.

\begin{acks}
This work is supported by the National Natural Science Foundation of China (62372399, 62476244), OPPO Research Fund, the Starry Night Science Fund of Zhejiang University Shanghai Institute for Advanced Study (SN-ZJU-SIAS-001), and the advanced computing resources provided by the Supercomputing Center of Hangzhou City University. 
We thank the reviewers for their valuable and insightful suggestions that improve the paper.
\end{acks}

\bibliographystyle{ACM-Reference-Format}
\bibliography{sample-base}


\begin{thebibliography}{81}


\ifx \showCODEN    \undefined \def \showCODEN     #1{\unskip}     \fi
\ifx \showISBNx    \undefined \def \showISBNx     #1{\unskip}     \fi
\ifx \showISBNxiii \undefined \def \showISBNxiii  #1{\unskip}     \fi
\ifx \showISSN     \undefined \def \showISSN      #1{\unskip}     \fi
\ifx \showLCCN     \undefined \def \showLCCN      #1{\unskip}     \fi
\ifx \shownote     \undefined \def \shownote      #1{#1}          \fi
\ifx \showarticletitle \undefined \def \showarticletitle #1{#1}   \fi
\ifx \showURL      \undefined \def \showURL       {\relax}        \fi
\providecommand\bibfield[2]{#2}
\providecommand\bibinfo[2]{#2}
\providecommand\natexlab[1]{#1}
\providecommand\showeprint[2][]{arXiv:#2}

\bibitem[Achiam et~al\mbox{.}(2023)]%
        {achiam2023gpt}
\bibfield{author}{\bibinfo{person}{Josh Achiam}, \bibinfo{person}{Steven Adler}, \bibinfo{person}{Sandhini Agarwal}, \bibinfo{person}{Lama Ahmad}, \bibinfo{person}{Ilge Akkaya}, \bibinfo{person}{Florencia~Leoni Aleman}, \bibinfo{person}{Diogo Almeida}, \bibinfo{person}{Janko Altenschmidt}, \bibinfo{person}{Sam Altman}, \bibinfo{person}{Shyamal Anadkat}, {et~al\mbox{.}}} \bibinfo{year}{2023}\natexlab{}.
\newblock \showarticletitle{Gpt-4 technical report}.
\newblock \bibinfo{journal}{\emph{arXiv preprint arXiv:2303.08774}} (\bibinfo{year}{2023}).
\newblock


\bibitem[Bao et~al\mbox{.}(2025)]%
        {bao2025bi}
\bibfield{author}{\bibinfo{person}{Keqin Bao}, \bibinfo{person}{Jizhi Zhang}, \bibinfo{person}{Wenjie Wang}, \bibinfo{person}{Yang Zhang}, \bibinfo{person}{Zhengyi Yang}, \bibinfo{person}{Yanchen Luo}, \bibinfo{person}{Chong Chen}, \bibinfo{person}{Fuli Feng}, {and} \bibinfo{person}{Qi Tian}.} \bibinfo{year}{2025}\natexlab{}.
\newblock \showarticletitle{A bi-step grounding paradigm for large language models in recommendation systems}.
\newblock \bibinfo{journal}{\emph{ACM Transactions on Recommender Systems}} \bibinfo{volume}{3}, \bibinfo{number}{4} (\bibinfo{year}{2025}), \bibinfo{pages}{1--27}.
\newblock


\bibitem[Bao et~al\mbox{.}(2024)]%
        {bao2024decoding}
\bibfield{author}{\bibinfo{person}{Keqin Bao}, \bibinfo{person}{Jizhi Zhang}, \bibinfo{person}{Yang Zhang}, \bibinfo{person}{Xinyue Huo}, \bibinfo{person}{Chong Chen}, {and} \bibinfo{person}{Fuli Feng}.} \bibinfo{year}{2024}\natexlab{}.
\newblock \showarticletitle{Decoding matters: Addressing amplification bias and homogeneity issue for llm-based recommendation}.
\newblock \bibinfo{journal}{\emph{arXiv preprint arXiv:2406.14900}} (\bibinfo{year}{2024}).
\newblock


\bibitem[Bao et~al\mbox{.}(2023)]%
        {bao2023tallrec}
\bibfield{author}{\bibinfo{person}{Keqin Bao}, \bibinfo{person}{Jizhi Zhang}, \bibinfo{person}{Yang Zhang}, \bibinfo{person}{Wenjie Wang}, \bibinfo{person}{Fuli Feng}, {and} \bibinfo{person}{Xiangnan He}.} \bibinfo{year}{2023}\natexlab{}.
\newblock \showarticletitle{Tallrec: An effective and efficient tuning framework to align large language model with recommendation}. In \bibinfo{booktitle}{\emph{Proceedings of the 17th ACM Conference on Recommender Systems}}. \bibinfo{pages}{1007--1014}.
\newblock


\bibitem[Barkan et~al\mbox{.}(2024)]%
        {barkan2024llm}
\bibfield{author}{\bibinfo{person}{Oren Barkan}, \bibinfo{person}{Yonatan Toib}, \bibinfo{person}{Yehonatan Elisha}, \bibinfo{person}{Jonathan Weill}, {and} \bibinfo{person}{Noam Koenigstein}.} \bibinfo{year}{2024}\natexlab{}.
\newblock \showarticletitle{LLM Explainability via Attributive Masking Learning}. In \bibinfo{booktitle}{\emph{Findings of the Association for Computational Linguistics: EMNLP 2024}}. \bibinfo{pages}{9522--9537}.
\newblock


\bibitem[Bito et~al\mbox{.}(2025)]%
        {bito2025evaluating}
\bibfield{author}{\bibinfo{person}{Ethan Bito}, \bibinfo{person}{Yongli Ren}, {and} \bibinfo{person}{Estrid He}.} \bibinfo{year}{2025}\natexlab{}.
\newblock \showarticletitle{Evaluating Position Bias in Large Language Model Recommendations}.
\newblock \bibinfo{journal}{\emph{arXiv preprint arXiv:2508.02020}} (\bibinfo{year}{2025}).
\newblock


\bibitem[Chen et~al\mbox{.}(2023)]%
        {chen2023bias}
\bibfield{author}{\bibinfo{person}{Jiawei Chen}, \bibinfo{person}{Hande Dong}, \bibinfo{person}{Xiang Wang}, \bibinfo{person}{Fuli Feng}, \bibinfo{person}{Meng Wang}, {and} \bibinfo{person}{Xiangnan He}.} \bibinfo{year}{2023}\natexlab{}.
\newblock \showarticletitle{Bias and debias in recommender system: A survey and future directions}.
\newblock \bibinfo{journal}{\emph{ACM Transactions on Information Systems}} \bibinfo{volume}{41}, \bibinfo{number}{3} (\bibinfo{year}{2023}), \bibinfo{pages}{1--39}.
\newblock


\bibitem[Chen et~al\mbox{.}(2025c)]%
        {Chen_2025_ICCV}
\bibfield{author}{\bibinfo{person}{Lu Chen}, \bibinfo{person}{Yizhou Wang}, \bibinfo{person}{Shixiang Tang}, \bibinfo{person}{Qianhong Ma}, \bibinfo{person}{Tong He}, \bibinfo{person}{Wanli Ouyang}, \bibinfo{person}{Xiaowei Zhou}, \bibinfo{person}{Hujun Bao}, {and} \bibinfo{person}{Sida Peng}.} \bibinfo{year}{2025}\natexlab{c}.
\newblock \showarticletitle{EgoAgent: A Joint Predictive Agent Model in Egocentric Worlds}. In \bibinfo{booktitle}{\emph{Proceedings of the IEEE/CVF International Conference on Computer Vision (ICCV)}}. \bibinfo{pages}{6970--6980}.
\newblock


\bibitem[Chen et~al\mbox{.}(2024a)]%
        {chen2024sigformer}
\bibfield{author}{\bibinfo{person}{Sirui Chen}, \bibinfo{person}{Jiawei Chen}, \bibinfo{person}{Sheng Zhou}, \bibinfo{person}{Bohao Wang}, \bibinfo{person}{Shen Han}, \bibinfo{person}{Chanfei Su}, \bibinfo{person}{Yuqing Yuan}, {and} \bibinfo{person}{Can Wang}.} \bibinfo{year}{2024}\natexlab{a}.
\newblock \showarticletitle{SIGformer: Sign-aware Graph Transformer for Recommendation}. In \bibinfo{booktitle}{\emph{Proceedings of the 47th International ACM SIGIR Conference on Research and Development in Information Retrieval}}. \bibinfo{pages}{1274--1284}.
\newblock


\bibitem[Chen et~al\mbox{.}(2025a)]%
        {chen2025rankformer}
\bibfield{author}{\bibinfo{person}{Sirui Chen}, \bibinfo{person}{Shen Han}, \bibinfo{person}{Jiawei Chen}, \bibinfo{person}{Binbin Hu}, \bibinfo{person}{Sheng Zhou}, \bibinfo{person}{Gang Wang}, \bibinfo{person}{Yan Feng}, \bibinfo{person}{Chun Chen}, {and} \bibinfo{person}{Can Wang}.} \bibinfo{year}{2025}\natexlab{a}.
\newblock \showarticletitle{Rankformer: A Graph Transformer for Recommendation based on Ranking Objective}. In \bibinfo{booktitle}{\emph{Proceedings of the ACM on Web Conference 2025}}. \bibinfo{pages}{3037--3048}.
\newblock


\bibitem[Chen et~al\mbox{.}(2025b)]%
        {chen2025arrows}
\bibfield{author}{\bibinfo{person}{Sirui Chen}, \bibinfo{person}{Changxin Tian}, \bibinfo{person}{Binbin Hu}, \bibinfo{person}{Kunlong Chen}, \bibinfo{person}{Ziqi Liu}, \bibinfo{person}{Zhiqiang Zhang}, {and} \bibinfo{person}{Jun Zhou}.} \bibinfo{year}{2025}\natexlab{b}.
\newblock \showarticletitle{Arrows of math reasoning data synthesis for large language models: Diversity, complexity and correctness}. In \bibinfo{booktitle}{\emph{Proceedings of the 34th ACM International Conference on Information and Knowledge Management}}. \bibinfo{pages}{4665--4669}.
\newblock


\bibitem[Chen et~al\mbox{.}(2024b)]%
        {chen2024softmax}
\bibfield{author}{\bibinfo{person}{Yuxin Chen}, \bibinfo{person}{Junfei Tan}, \bibinfo{person}{An Zhang}, \bibinfo{person}{Zhengyi Yang}, \bibinfo{person}{Leheng Sheng}, \bibinfo{person}{Enzhi Zhang}, \bibinfo{person}{Xiang Wang}, {and} \bibinfo{person}{Tat-Seng Chua}.} \bibinfo{year}{2024}\natexlab{b}.
\newblock \showarticletitle{On Softmax Direct Preference Optimization for Recommendation}.
\newblock \bibinfo{journal}{\emph{arXiv preprint arXiv:2406.09215}} (\bibinfo{year}{2024}).
\newblock


\bibitem[Creager et~al\mbox{.}(2021)]%
        {creager2021environment}
\bibfield{author}{\bibinfo{person}{Elliot Creager}, \bibinfo{person}{J{\"o}rn-Henrik Jacobsen}, {and} \bibinfo{person}{Richard Zemel}.} \bibinfo{year}{2021}\natexlab{}.
\newblock \showarticletitle{Environment inference for invariant learning}. In \bibinfo{booktitle}{\emph{International Conference on Machine Learning}}. PMLR, \bibinfo{pages}{2189--2200}.
\newblock


\bibitem[Cui et~al\mbox{.}(2025a)]%
        {cui2025hatllm}
\bibfield{author}{\bibinfo{person}{Yu Cui}, \bibinfo{person}{Feng Liu}, \bibinfo{person}{Jiawei Chen}, \bibinfo{person}{Canghong Jin}, \bibinfo{person}{Xingyu Lou}, \bibinfo{person}{Changwang Zhang}, \bibinfo{person}{Jun Wang}, \bibinfo{person}{Yuegang Sun}, {and} \bibinfo{person}{Can Wang}.} \bibinfo{year}{2025}\natexlab{a}.
\newblock \showarticletitle{HatLLM: Hierarchical Attention Masking for Enhanced Collaborative Modeling in LLM-based Recommendation}.
\newblock \bibinfo{journal}{\emph{arXiv preprint arXiv:2510.10955}} (\bibinfo{year}{2025}).
\newblock


\bibitem[Cui et~al\mbox{.}(2025b)]%
        {cui2025field}
\bibfield{author}{\bibinfo{person}{Yu Cui}, \bibinfo{person}{Feng Liu}, \bibinfo{person}{Jiawei Chen}, \bibinfo{person}{Xingyu Lou}, \bibinfo{person}{Changwang Zhang}, \bibinfo{person}{Jun Wang}, \bibinfo{person}{Yuegang Sun}, \bibinfo{person}{Xiaohu Yang}, {and} \bibinfo{person}{Can Wang}.} \bibinfo{year}{2025}\natexlab{b}.
\newblock \showarticletitle{Field Matters: A lightweight LLM-enhanced Method for CTR Prediction}.
\newblock \bibinfo{journal}{\emph{arXiv preprint arXiv:2505.14057}} (\bibinfo{year}{2025}).
\newblock


\bibitem[Cui et~al\mbox{.}(2024)]%
        {cui2024distillation}
\bibfield{author}{\bibinfo{person}{Yu Cui}, \bibinfo{person}{Feng Liu}, \bibinfo{person}{Pengbo Wang}, \bibinfo{person}{Bohao Wang}, \bibinfo{person}{Heng Tang}, \bibinfo{person}{Yi Wan}, \bibinfo{person}{Jun Wang}, {and} \bibinfo{person}{Jiawei Chen}.} \bibinfo{year}{2024}\natexlab{}.
\newblock \showarticletitle{Distillation Matters: Empowering Sequential Recommenders to Match the Performance of Large Language Models}. In \bibinfo{booktitle}{\emph{Proceedings of the 18th ACM Conference on Recommender Systems}}. \bibinfo{pages}{507--517}.
\newblock


\bibitem[Dai et~al\mbox{.}(2024)]%
        {dai2024bias}
\bibfield{author}{\bibinfo{person}{Sunhao Dai}, \bibinfo{person}{Chen Xu}, \bibinfo{person}{Shicheng Xu}, \bibinfo{person}{Liang Pang}, \bibinfo{person}{Zhenhua Dong}, {and} \bibinfo{person}{Jun Xu}.} \bibinfo{year}{2024}\natexlab{}.
\newblock \showarticletitle{Bias and unfairness in information retrieval systems: New challenges in the llm era}. In \bibinfo{booktitle}{\emph{Proceedings of the 30th ACM SIGKDD Conference on Knowledge Discovery and Data Mining}}. \bibinfo{pages}{6437--6447}.
\newblock


\bibitem[Deldjoo(2024)]%
        {deldjoo2024understanding}
\bibfield{author}{\bibinfo{person}{Yashar Deldjoo}.} \bibinfo{year}{2024}\natexlab{}.
\newblock \showarticletitle{Understanding biases in ChatGPT-based recommender systems: Provider fairness, temporal stability, and recency}.
\newblock \bibinfo{journal}{\emph{ACM Transactions on Recommender Systems}} (\bibinfo{year}{2024}).
\newblock


\bibitem[Dubey et~al\mbox{.}(2024)]%
        {dubey2024llama}
\bibfield{author}{\bibinfo{person}{Abhimanyu Dubey}, \bibinfo{person}{Abhinav Jauhri}, \bibinfo{person}{Abhinav Pandey}, \bibinfo{person}{Abhishek Kadian}, \bibinfo{person}{Ahmad Al-Dahle}, \bibinfo{person}{Aiesha Letman}, \bibinfo{person}{Akhil Mathur}, \bibinfo{person}{Alan Schelten}, \bibinfo{person}{Amy Yang}, \bibinfo{person}{Angela Fan}, {et~al\mbox{.}}} \bibinfo{year}{2024}\natexlab{}.
\newblock \showarticletitle{The Llama 3 Herd of Models}.
\newblock \bibinfo{journal}{\emph{arXiv preprint arXiv:2407.21783}} (\bibinfo{year}{2024}).
\newblock


\bibitem[Fang et~al\mbox{.}(2020)]%
        {fang2020deep}
\bibfield{author}{\bibinfo{person}{Hui Fang}, \bibinfo{person}{Danning Zhang}, \bibinfo{person}{Yiheng Shu}, {and} \bibinfo{person}{Guibing Guo}.} \bibinfo{year}{2020}\natexlab{}.
\newblock \showarticletitle{Deep learning for sequential recommendation: Algorithms, influential factors, and evaluations}.
\newblock \bibinfo{journal}{\emph{ACM Transactions on Information Systems (TOIS)}} \bibinfo{volume}{39}, \bibinfo{number}{1} (\bibinfo{year}{2020}), \bibinfo{pages}{1--42}.
\newblock


\bibitem[Gao et~al\mbox{.}(2025a)]%
        {gao2025sprec}
\bibfield{author}{\bibinfo{person}{Chongming Gao}, \bibinfo{person}{Ruijun Chen}, \bibinfo{person}{Shuai Yuan}, \bibinfo{person}{Kexin Huang}, \bibinfo{person}{Yuanqing Yu}, {and} \bibinfo{person}{Xiangnan He}.} \bibinfo{year}{2025}\natexlab{a}.
\newblock \showarticletitle{Sprec: Self-play to debias llm-based recommendation}. In \bibinfo{booktitle}{\emph{Proceedings of the ACM on Web Conference 2025}}. \bibinfo{pages}{5075--5084}.
\newblock


\bibitem[Gao et~al\mbox{.}(2025b)]%
        {gao2025process}
\bibfield{author}{\bibinfo{person}{Chongming Gao}, \bibinfo{person}{Mengyao Gao}, \bibinfo{person}{Chenxiao Fan}, \bibinfo{person}{Shuai Yuan}, \bibinfo{person}{Wentao Shi}, {and} \bibinfo{person}{Xiangnan He}.} \bibinfo{year}{2025}\natexlab{b}.
\newblock \showarticletitle{Process-supervised llm recommenders via flow-guided tuning}. In \bibinfo{booktitle}{\emph{Proceedings of the 48th International ACM SIGIR Conference on Research and Development in Information Retrieval}}. \bibinfo{pages}{1934--1943}.
\newblock


\bibitem[Guo et~al\mbox{.}(2025)]%
        {guo2025deepseek}
\bibfield{author}{\bibinfo{person}{Daya Guo}, \bibinfo{person}{Dejian Yang}, \bibinfo{person}{Haowei Zhang}, \bibinfo{person}{Junxiao Song}, \bibinfo{person}{Peiyi Wang}, \bibinfo{person}{Qihao Zhu}, \bibinfo{person}{Runxin Xu}, \bibinfo{person}{Ruoyu Zhang}, \bibinfo{person}{Shirong Ma}, \bibinfo{person}{Xiao Bi}, {et~al\mbox{.}}} \bibinfo{year}{2025}\natexlab{}.
\newblock \showarticletitle{DeepSeek-R1 incentivizes reasoning in LLMs through reinforcement learning}.
\newblock \bibinfo{journal}{\emph{Nature}} \bibinfo{volume}{645}, \bibinfo{number}{8081} (\bibinfo{year}{2025}), \bibinfo{pages}{633--638}.
\newblock


\bibitem[Han et~al\mbox{.}(2024)]%
        {han2024mitigating}
\bibfield{author}{\bibinfo{person}{Hyeonggeun Han}, \bibinfo{person}{Sehwan Kim}, \bibinfo{person}{Hyungjun Joo}, \bibinfo{person}{Sangwoo Hong}, {and} \bibinfo{person}{Jungwoo Lee}.} \bibinfo{year}{2024}\natexlab{}.
\newblock \showarticletitle{Mitigating spurious correlations via disagreement probability}.
\newblock \bibinfo{journal}{\emph{Advances in Neural Information Processing Systems}}  \bibinfo{volume}{37} (\bibinfo{year}{2024}), \bibinfo{pages}{74363--74382}.
\newblock


\bibitem[Hashimoto et~al\mbox{.}(2018)]%
        {hashimoto2018fairness}
\bibfield{author}{\bibinfo{person}{Tatsunori Hashimoto}, \bibinfo{person}{Megha Srivastava}, \bibinfo{person}{Hongseok Namkoong}, {and} \bibinfo{person}{Percy Liang}.} \bibinfo{year}{2018}\natexlab{}.
\newblock \showarticletitle{Fairness without demographics in repeated loss minimization}. In \bibinfo{booktitle}{\emph{International Conference on Machine Learning}}. PMLR, \bibinfo{pages}{1929--1938}.
\newblock


\bibitem[Hidasi et~al\mbox{.}(2015)]%
        {hidasi2015session}
\bibfield{author}{\bibinfo{person}{Bal{\'a}zs Hidasi}, \bibinfo{person}{Alexandros Karatzoglou}, \bibinfo{person}{Linas Baltrunas}, {and} \bibinfo{person}{Domonkos Tikk}.} \bibinfo{year}{2015}\natexlab{}.
\newblock \showarticletitle{Session-based recommendations with recurrent neural networks}.
\newblock \bibinfo{journal}{\emph{arXiv preprint arXiv:1511.06939}} (\bibinfo{year}{2015}).
\newblock


\bibitem[Hou et~al\mbox{.}(2024)]%
        {hou2024large}
\bibfield{author}{\bibinfo{person}{Yupeng Hou}, \bibinfo{person}{Junjie Zhang}, \bibinfo{person}{Zihan Lin}, \bibinfo{person}{Hongyu Lu}, \bibinfo{person}{Ruobing Xie}, \bibinfo{person}{Julian McAuley}, {and} \bibinfo{person}{Wayne~Xin Zhao}.} \bibinfo{year}{2024}\natexlab{}.
\newblock \showarticletitle{Large language models are zero-shot rankers for recommender systems}. In \bibinfo{booktitle}{\emph{European Conference on Information Retrieval}}. Springer, \bibinfo{pages}{364--381}.
\newblock


\bibitem[Jiang et~al\mbox{.}(2025)]%
        {jiang2025beyond}
\bibfield{author}{\bibinfo{person}{Chumeng Jiang}, \bibinfo{person}{Jiayin Wang}, \bibinfo{person}{Weizhi Ma}, \bibinfo{person}{Charles~LA Clarke}, \bibinfo{person}{Shuai Wang}, \bibinfo{person}{Chuhan Wu}, {and} \bibinfo{person}{Min Zhang}.} \bibinfo{year}{2025}\natexlab{}.
\newblock \showarticletitle{Beyond Utility: Evaluating LLM as Recommender}. In \bibinfo{booktitle}{\emph{Proceedings of the ACM on Web Conference 2025}}. \bibinfo{pages}{3850--3862}.
\newblock


\bibitem[Jiang et~al\mbox{.}(2024)]%
        {jiang2024item}
\bibfield{author}{\bibinfo{person}{Meng Jiang}, \bibinfo{person}{Keqin Bao}, \bibinfo{person}{Jizhi Zhang}, \bibinfo{person}{Wenjie Wang}, \bibinfo{person}{Zhengyi Yang}, \bibinfo{person}{Fuli Feng}, {and} \bibinfo{person}{Xiangnan He}.} \bibinfo{year}{2024}\natexlab{}.
\newblock \showarticletitle{Item-side fairness of large language model-based recommendation system}. In \bibinfo{booktitle}{\emph{Proceedings of the ACM Web Conference 2024}}. \bibinfo{pages}{4717--4726}.
\newblock


\bibitem[Kang and McAuley(2018)]%
        {kang2018self}
\bibfield{author}{\bibinfo{person}{Wang-Cheng Kang} {and} \bibinfo{person}{Julian McAuley}.} \bibinfo{year}{2018}\natexlab{}.
\newblock \showarticletitle{Self-attentive sequential recommendation}. In \bibinfo{booktitle}{\emph{2018 IEEE international conference on data mining (ICDM)}}. IEEE, \bibinfo{pages}{197--206}.
\newblock


\bibitem[Kauf et~al\mbox{.}(2024)]%
        {kauf2024log}
\bibfield{author}{\bibinfo{person}{Carina Kauf}, \bibinfo{person}{Emmanuele Chersoni}, \bibinfo{person}{Alessandro Lenci}, \bibinfo{person}{Evelina Fedorenko}, {and} \bibinfo{person}{Anna Ivanova}.} \bibinfo{year}{2024}\natexlab{}.
\newblock \showarticletitle{Log Probabilities Are a Reliable Estimate of Semantic Plausibility in Base and Instruction-Tuned Language Models}. In \bibinfo{booktitle}{\emph{Proceedings of the 7th BlackboxNLP Workshop: Analyzing and Interpreting Neural Networks for NLP}}. \bibinfo{pages}{263--277}.
\newblock


\bibitem[Kim et~al\mbox{.}(2024)]%
        {kim2024large}
\bibfield{author}{\bibinfo{person}{Sein Kim}, \bibinfo{person}{Hongseok Kang}, \bibinfo{person}{Seungyoon Choi}, \bibinfo{person}{Donghyun Kim}, \bibinfo{person}{Minchul Yang}, {and} \bibinfo{person}{Chanyoung Park}.} \bibinfo{year}{2024}\natexlab{}.
\newblock \showarticletitle{Large language models meet collaborative filtering: An efficient all-round llm-based recommender system}. In \bibinfo{booktitle}{\emph{Proceedings of the 30th ACM SIGKDD Conference on Knowledge Discovery and Data Mining}}. \bibinfo{pages}{1395--1406}.
\newblock


\bibitem[Kokhlikyan et~al\mbox{.}(2020)]%
        {kokhlikyan2020captum}
\bibfield{author}{\bibinfo{person}{Narine Kokhlikyan}, \bibinfo{person}{Vivek Miglani}, \bibinfo{person}{Miguel Martin}, \bibinfo{person}{Edward Wang}, \bibinfo{person}{Bilal Alsallakh}, \bibinfo{person}{Jonathan Reynolds}, \bibinfo{person}{Alexander Melnikov}, \bibinfo{person}{Natalia Kliushkina}, \bibinfo{person}{Carlos Araya}, \bibinfo{person}{Siqi Yan}, {et~al\mbox{.}}} \bibinfo{year}{2020}\natexlab{}.
\newblock \showarticletitle{Captum: A unified and generic model interpretability library for pytorch}.
\newblock \bibinfo{journal}{\emph{arXiv preprint arXiv:2009.07896}} (\bibinfo{year}{2020}).
\newblock


\bibitem[Lepage et~al\mbox{.}(2025)]%
        {lepage2025closing}
\bibfield{author}{\bibinfo{person}{Simon Lepage}, \bibinfo{person}{Jeremie Mary}, {and} \bibinfo{person}{David Picard}.} \bibinfo{year}{2025}\natexlab{}.
\newblock \showarticletitle{Closing the Performance Gap in Generative Recommenders with Collaborative Tokenization and Efficient Modeling}.
\newblock \bibinfo{journal}{\emph{arXiv preprint arXiv:2508.14910}} (\bibinfo{year}{2025}).
\newblock


\bibitem[Li et~al\mbox{.}(2023b)]%
        {li2023large}
\bibfield{author}{\bibinfo{person}{Lei Li}, \bibinfo{person}{Yongfeng Zhang}, \bibinfo{person}{Dugang Liu}, {and} \bibinfo{person}{Li Chen}.} \bibinfo{year}{2023}\natexlab{b}.
\newblock \showarticletitle{Large language models for generative recommendation: A survey and visionary discussions}.
\newblock \bibinfo{journal}{\emph{arXiv preprint arXiv:2309.01157}} (\bibinfo{year}{2023}).
\newblock


\bibitem[Li et~al\mbox{.}(2023a)]%
        {li2023preliminary}
\bibfield{author}{\bibinfo{person}{Xinyi Li}, \bibinfo{person}{Yongfeng Zhang}, {and} \bibinfo{person}{Edward~C Malthouse}.} \bibinfo{year}{2023}\natexlab{a}.
\newblock \showarticletitle{A preliminary study of chatgpt on news recommendation: Personalization, provider fairness, fake news}.
\newblock \bibinfo{journal}{\emph{arXiv preprint arXiv:2306.10702}} (\bibinfo{year}{2023}).
\newblock


\bibitem[Liao et~al\mbox{.}(2024a)]%
        {liao2024rosepo}
\bibfield{author}{\bibinfo{person}{Jiayi Liao}, \bibinfo{person}{Xiangnan He}, \bibinfo{person}{Ruobing Xie}, \bibinfo{person}{Jiancan Wu}, \bibinfo{person}{Yancheng Yuan}, \bibinfo{person}{Xingwu Sun}, \bibinfo{person}{Zhanhui Kang}, {and} \bibinfo{person}{Xiang Wang}.} \bibinfo{year}{2024}\natexlab{a}.
\newblock \showarticletitle{RosePO: Aligning LLM-based Recommenders with Human Values}.
\newblock \bibinfo{journal}{\emph{arXiv preprint arXiv:2410.12519}} (\bibinfo{year}{2024}).
\newblock


\bibitem[Liao et~al\mbox{.}(2024b)]%
        {liao2024llara}
\bibfield{author}{\bibinfo{person}{Jiayi Liao}, \bibinfo{person}{Sihang Li}, \bibinfo{person}{Zhengyi Yang}, \bibinfo{person}{Jiancan Wu}, \bibinfo{person}{Yancheng Yuan}, \bibinfo{person}{Xiang Wang}, {and} \bibinfo{person}{Xiangnan He}.} \bibinfo{year}{2024}\natexlab{b}.
\newblock \showarticletitle{Llara: Large language-recommendation assistant}. In \bibinfo{booktitle}{\emph{Proceedings of the 47th International ACM SIGIR Conference on Research and Development in Information Retrieval}}. \bibinfo{pages}{1785--1795}.
\newblock


\bibitem[Lin et~al\mbox{.}(2024)]%
        {lin2024rella}
\bibfield{author}{\bibinfo{person}{Jianghao Lin}, \bibinfo{person}{Rong Shan}, \bibinfo{person}{Chenxu Zhu}, \bibinfo{person}{Kounianhua Du}, \bibinfo{person}{Bo Chen}, \bibinfo{person}{Shigang Quan}, \bibinfo{person}{Ruiming Tang}, \bibinfo{person}{Yong Yu}, {and} \bibinfo{person}{Weinan Zhang}.} \bibinfo{year}{2024}\natexlab{}.
\newblock \showarticletitle{Rella: Retrieval-enhanced large language models for lifelong sequential behavior comprehension in recommendation}. In \bibinfo{booktitle}{\emph{Proceedings of the ACM on Web Conference 2024}}. \bibinfo{pages}{3497--3508}.
\newblock


\bibitem[Lin et~al\mbox{.}(2025)]%
        {lin2025recommendation}
\bibfield{author}{\bibinfo{person}{Siyi Lin}, \bibinfo{person}{Chongming Gao}, \bibinfo{person}{Jiawei Chen}, \bibinfo{person}{Sheng Zhou}, \bibinfo{person}{Binbin Hu}, \bibinfo{person}{Yan Feng}, \bibinfo{person}{Chun Chen}, {and} \bibinfo{person}{Can Wang}.} \bibinfo{year}{2025}\natexlab{}.
\newblock \showarticletitle{How do recommendation models amplify popularity bias? An analysis from the spectral perspective}. In \bibinfo{booktitle}{\emph{Proceedings of the Eighteenth ACM International Conference on Web Search and Data Mining}}. \bibinfo{pages}{659--668}.
\newblock


\bibitem[Lu et~al\mbox{.}(2025)]%
        {lu2025dual}
\bibfield{author}{\bibinfo{person}{Sijin Lu}, \bibinfo{person}{Zhibo Man}, \bibinfo{person}{Fangyuan Luo}, {and} \bibinfo{person}{Jun Wu}.} \bibinfo{year}{2025}\natexlab{}.
\newblock \showarticletitle{Dual Debiasing in LLM-based Recommendation}. In \bibinfo{booktitle}{\emph{Proceedings of the 48th International ACM SIGIR Conference on Research and Development in Information Retrieval}}. \bibinfo{pages}{2685--2689}.
\newblock


\bibitem[Luo et~al\mbox{.}(2025)]%
        {luo2025recranker}
\bibfield{author}{\bibinfo{person}{Sichun Luo}, \bibinfo{person}{Bowei He}, \bibinfo{person}{Haohan Zhao}, \bibinfo{person}{Wei Shao}, \bibinfo{person}{Yanlin Qi}, \bibinfo{person}{Yinya Huang}, \bibinfo{person}{Aojun Zhou}, \bibinfo{person}{Yuxuan Yao}, \bibinfo{person}{Zongpeng Li}, \bibinfo{person}{Yuanzhang Xiao}, {et~al\mbox{.}}} \bibinfo{year}{2025}\natexlab{}.
\newblock \showarticletitle{Recranker: Instruction tuning large language model as ranker for top-k recommendation}.
\newblock \bibinfo{journal}{\emph{ACM Transactions on Information Systems}} \bibinfo{volume}{43}, \bibinfo{number}{5} (\bibinfo{year}{2025}), \bibinfo{pages}{1--31}.
\newblock


\bibitem[Ma et~al\mbox{.}(2023)]%
        {ma2023large}
\bibfield{author}{\bibinfo{person}{Tianhui Ma}, \bibinfo{person}{Yuan Cheng}, \bibinfo{person}{Hengshu Zhu}, {and} \bibinfo{person}{Hui Xiong}.} \bibinfo{year}{2023}\natexlab{}.
\newblock \showarticletitle{Large language models are not stable recommender systems}.
\newblock \bibinfo{journal}{\emph{arXiv preprint arXiv:2312.15746}} (\bibinfo{year}{2023}).
\newblock


\bibitem[Miglani et~al\mbox{.}(2023)]%
        {miglani2023using}
\bibfield{author}{\bibinfo{person}{Vivek Miglani}, \bibinfo{person}{Aobo Yang}, \bibinfo{person}{Aram~H Markosyan}, \bibinfo{person}{Diego Garcia-Olano}, {and} \bibinfo{person}{Narine Kokhlikyan}.} \bibinfo{year}{2023}\natexlab{}.
\newblock \showarticletitle{Using captum to explain generative language models}.
\newblock \bibinfo{journal}{\emph{arXiv preprint arXiv:2312.05491}} (\bibinfo{year}{2023}).
\newblock


\bibitem[Na et~al\mbox{.}(2024)]%
        {na2024enhancing}
\bibfield{author}{\bibinfo{person}{Hyunsoo Na}, \bibinfo{person}{Minseok Gang}, \bibinfo{person}{Youngrok Ko}, \bibinfo{person}{Jinseok Seol}, {and} \bibinfo{person}{Sang-goo Lee}.} \bibinfo{year}{2024}\natexlab{}.
\newblock \showarticletitle{Enhancing Large Language Model Based Sequential Recommender Systems with Pseudo Labels Reconstruction}. In \bibinfo{booktitle}{\emph{Findings of the Association for Computational Linguistics: EMNLP 2024}}. \bibinfo{pages}{7213--7222}.
\newblock


\bibitem[Oren et~al\mbox{.}(2019)]%
        {oren2019distributionally}
\bibfield{author}{\bibinfo{person}{Yonatan Oren}, \bibinfo{person}{Shiori Sagawa}, \bibinfo{person}{Tatsunori~B Hashimoto}, {and} \bibinfo{person}{Percy Liang}.} \bibinfo{year}{2019}\natexlab{}.
\newblock \showarticletitle{Distributionally Robust Language Modeling}. In \bibinfo{booktitle}{\emph{Proceedings of the 2019 Conference on Empirical Methods in Natural Language Processing and the 9th International Joint Conference on Natural Language Processing (EMNLP-IJCNLP)}}. \bibinfo{pages}{4227--4237}.
\newblock


\bibitem[Ouyang et~al\mbox{.}(2022)]%
        {ouyang2022training}
\bibfield{author}{\bibinfo{person}{Long Ouyang}, \bibinfo{person}{Jeffrey Wu}, \bibinfo{person}{Xu Jiang}, \bibinfo{person}{Diogo Almeida}, \bibinfo{person}{Carroll Wainwright}, \bibinfo{person}{Pamela Mishkin}, \bibinfo{person}{Chong Zhang}, \bibinfo{person}{Sandhini Agarwal}, \bibinfo{person}{Katarina Slama}, \bibinfo{person}{Alex Ray}, {et~al\mbox{.}}} \bibinfo{year}{2022}\natexlab{}.
\newblock \showarticletitle{Training language models to follow instructions with human feedback}.
\newblock \bibinfo{journal}{\emph{Advances in neural information processing systems}}  \bibinfo{volume}{35} (\bibinfo{year}{2022}), \bibinfo{pages}{27730--27744}.
\newblock


\bibitem[Rafailov et~al\mbox{.}(2024)]%
        {rafailov2024direct}
\bibfield{author}{\bibinfo{person}{Rafael Rafailov}, \bibinfo{person}{Archit Sharma}, \bibinfo{person}{Eric Mitchell}, \bibinfo{person}{Christopher~D Manning}, \bibinfo{person}{Stefano Ermon}, {and} \bibinfo{person}{Chelsea Finn}.} \bibinfo{year}{2024}\natexlab{}.
\newblock \showarticletitle{Direct preference optimization: Your language model is secretly a reward model}.
\newblock \bibinfo{journal}{\emph{Advances in Neural Information Processing Systems}}  \bibinfo{volume}{36} (\bibinfo{year}{2024}).
\newblock


\bibitem[Rahimian and Mehrotra(2019)]%
        {rahimian2019distributionally}
\bibfield{author}{\bibinfo{person}{Hamed Rahimian} {and} \bibinfo{person}{Sanjay Mehrotra}.} \bibinfo{year}{2019}\natexlab{}.
\newblock \showarticletitle{Distributionally robust optimization: A review}.
\newblock \bibinfo{journal}{\emph{arXiv preprint arXiv:1908.05659}} (\bibinfo{year}{2019}).
\newblock


\bibitem[Sagawa et~al\mbox{.}(2019)]%
        {sagawa2019distributionally}
\bibfield{author}{\bibinfo{person}{Shiori Sagawa}, \bibinfo{person}{Pang~Wei Koh}, \bibinfo{person}{Tatsunori~B Hashimoto}, {and} \bibinfo{person}{Percy Liang}.} \bibinfo{year}{2019}\natexlab{}.
\newblock \showarticletitle{Distributionally robust neural networks for group shifts: On the importance of regularization for worst-case generalization}.
\newblock \bibinfo{journal}{\emph{arXiv preprint arXiv:1911.08731}} (\bibinfo{year}{2019}).
\newblock


\bibitem[Sun et~al\mbox{.}(2019)]%
        {sun2019bert4rec}
\bibfield{author}{\bibinfo{person}{Fei Sun}, \bibinfo{person}{Jun Liu}, \bibinfo{person}{Jian Wu}, \bibinfo{person}{Changhua Pei}, \bibinfo{person}{Xiao Lin}, \bibinfo{person}{Wenwu Ou}, {and} \bibinfo{person}{Peng Jiang}.} \bibinfo{year}{2019}\natexlab{}.
\newblock \showarticletitle{BERT4Rec: Sequential recommendation with bidirectional encoder representations from transformer}. In \bibinfo{booktitle}{\emph{Proceedings of the 28th ACM international conference on information and knowledge management}}. \bibinfo{pages}{1441--1450}.
\newblock


\bibitem[Tan et~al\mbox{.}(2024)]%
        {tan2024idgenrec}
\bibfield{author}{\bibinfo{person}{Juntao Tan}, \bibinfo{person}{Shuyuan Xu}, \bibinfo{person}{Wenyue Hua}, \bibinfo{person}{Yingqiang Ge}, \bibinfo{person}{Zelong Li}, {and} \bibinfo{person}{Yongfeng Zhang}.} \bibinfo{year}{2024}\natexlab{}.
\newblock \showarticletitle{Idgenrec: Llm-recsys alignment with textual id learning}. In \bibinfo{booktitle}{\emph{Proceedings of the 47th international ACM SIGIR conference on research and development in information retrieval}}. \bibinfo{pages}{355--364}.
\newblock


\bibitem[Tang and Wang(2018)]%
        {tang2018personalized}
\bibfield{author}{\bibinfo{person}{Jiaxi Tang} {and} \bibinfo{person}{Ke Wang}.} \bibinfo{year}{2018}\natexlab{}.
\newblock \showarticletitle{Personalized top-n sequential recommendation via convolutional sequence embedding}. In \bibinfo{booktitle}{\emph{Proceedings of the eleventh ACM international conference on web search and data mining}}. \bibinfo{pages}{565--573}.
\newblock


\bibitem[Tang et~al\mbox{.}(2024)]%
        {tang2024graphgpt}
\bibfield{author}{\bibinfo{person}{Jiabin Tang}, \bibinfo{person}{Yuhao Yang}, \bibinfo{person}{Wei Wei}, \bibinfo{person}{Lei Shi}, \bibinfo{person}{Lixin Su}, \bibinfo{person}{Suqi Cheng}, \bibinfo{person}{Dawei Yin}, {and} \bibinfo{person}{Chao Huang}.} \bibinfo{year}{2024}\natexlab{}.
\newblock \showarticletitle{Graphgpt: Graph instruction tuning for large language models}. In \bibinfo{booktitle}{\emph{Proceedings of the 47th International ACM SIGIR Conference on Research and Development in Information Retrieval}}. \bibinfo{pages}{491--500}.
\newblock


\bibitem[Team et~al\mbox{.}(2024)]%
        {team2024qwen2}
\bibfield{author}{\bibinfo{person}{Qwen Team} {et~al\mbox{.}}} \bibinfo{year}{2024}\natexlab{}.
\newblock \showarticletitle{Qwen2 technical report}.
\newblock \bibinfo{journal}{\emph{arXiv preprint arXiv:2407.10671}} \bibinfo{volume}{2}, \bibinfo{number}{3} (\bibinfo{year}{2024}).
\newblock


\bibitem[Vaswani(2017)]%
        {vaswani2017attention}
\bibfield{author}{\bibinfo{person}{Ashish Vaswani}.} \bibinfo{year}{2017}\natexlab{}.
\newblock \showarticletitle{Attention is all you need}.
\newblock \bibinfo{journal}{\emph{arXiv preprint arXiv:1706.03762}} (\bibinfo{year}{2017}).
\newblock


\bibitem[Wang et~al\mbox{.}(2024a)]%
        {wang2024distributionally}
\bibfield{author}{\bibinfo{person}{Bohao Wang}, \bibinfo{person}{Jiawei Chen}, \bibinfo{person}{Changdong Li}, \bibinfo{person}{Sheng Zhou}, \bibinfo{person}{Qihao Shi}, \bibinfo{person}{Yang Gao}, \bibinfo{person}{Yan Feng}, \bibinfo{person}{Chun Chen}, {and} \bibinfo{person}{Can Wang}.} \bibinfo{year}{2024}\natexlab{a}.
\newblock \showarticletitle{Distributionally Robust Graph-based Recommendation System}. In \bibinfo{booktitle}{\emph{Proceedings of the ACM on Web Conference 2024}}. \bibinfo{pages}{3777--3788}.
\newblock


\bibitem[Wang et~al\mbox{.}(2025b)]%
        {wang2025msl}
\bibfield{author}{\bibinfo{person}{Bohao Wang}, \bibinfo{person}{Feng Liu}, \bibinfo{person}{Jiawei Chen}, \bibinfo{person}{Xingyu Lou}, \bibinfo{person}{Changwang Zhang}, \bibinfo{person}{Jun Wang}, \bibinfo{person}{Yuegang Sun}, \bibinfo{person}{Yan Feng}, \bibinfo{person}{Chun Chen}, {and} \bibinfo{person}{Can Wang}.} \bibinfo{year}{2025}\natexlab{b}.
\newblock \showarticletitle{Msl: Not all tokens are what you need for tuning llm as a recommender}. In \bibinfo{booktitle}{\emph{Proceedings of the 48th International ACM SIGIR Conference on Research and Development in Information Retrieval}}. \bibinfo{pages}{1912--1922}.
\newblock


\bibitem[Wang et~al\mbox{.}(2025c)]%
        {wang2025llm4dsr}
\bibfield{author}{\bibinfo{person}{Bohao Wang}, \bibinfo{person}{Feng Liu}, \bibinfo{person}{Changwang Zhang}, \bibinfo{person}{Jiawei Chen}, \bibinfo{person}{Yudi Wu}, \bibinfo{person}{Sheng Zhou}, \bibinfo{person}{Xingyu Lou}, \bibinfo{person}{Jun Wang}, \bibinfo{person}{Yan Feng}, \bibinfo{person}{Chun Chen}, {et~al\mbox{.}}} \bibinfo{year}{2025}\natexlab{c}.
\newblock \showarticletitle{Llm4dsr: Leveraging large language model for denoising sequential recommendation}.
\newblock \bibinfo{journal}{\emph{ACM Transactions on Information Systems}} \bibinfo{volume}{44}, \bibinfo{number}{1} (\bibinfo{year}{2025}), \bibinfo{pages}{1--32}.
\newblock


\bibitem[Wang et~al\mbox{.}(2024b)]%
        {wang2024flip}
\bibfield{author}{\bibinfo{person}{Hangyu Wang}, \bibinfo{person}{Jianghao Lin}, \bibinfo{person}{Xiangyang Li}, \bibinfo{person}{Bo Chen}, \bibinfo{person}{Chenxu Zhu}, \bibinfo{person}{Ruiming Tang}, \bibinfo{person}{Weinan Zhang}, {and} \bibinfo{person}{Yong Yu}.} \bibinfo{year}{2024}\natexlab{b}.
\newblock \showarticletitle{Flip: Fine-grained alignment between id-based models and pretrained language models for ctr prediction}. In \bibinfo{booktitle}{\emph{Proceedings of the 18th ACM conference on recommender systems}}. \bibinfo{pages}{94--104}.
\newblock


\bibitem[Wang and Lim(2023)]%
        {wang2023zero}
\bibfield{author}{\bibinfo{person}{Lei Wang} {and} \bibinfo{person}{Ee-Peng Lim}.} \bibinfo{year}{2023}\natexlab{}.
\newblock \showarticletitle{Zero-shot next-item recommendation using large pretrained language models}.
\newblock \bibinfo{journal}{\emph{arXiv preprint arXiv:2304.03153}} (\bibinfo{year}{2023}).
\newblock


\bibitem[Wang et~al\mbox{.}(2025a)]%
        {wang2025adaptive}
\bibfield{author}{\bibinfo{person}{Yu Wang}, \bibinfo{person}{Junshu Dai}, \bibinfo{person}{Yuchen Ying}, \bibinfo{person}{Yuxuan Liang}, \bibinfo{person}{Tongya Zheng}, {and} \bibinfo{person}{Mingli Song}.} \bibinfo{year}{2025}\natexlab{a}.
\newblock \showarticletitle{Adaptive Location Hierarchy Learning for Long-Tailed Mobility Prediction}.
\newblock \bibinfo{journal}{\emph{arXiv preprint arXiv:2505.19965}} (\bibinfo{year}{2025}).
\newblock


\bibitem[Wang et~al\mbox{.}(2024c)]%
        {wang2024cola}
\bibfield{author}{\bibinfo{person}{Yu Wang}, \bibinfo{person}{Tongya Zheng}, \bibinfo{person}{Yuxuan Liang}, \bibinfo{person}{Shunyu Liu}, {and} \bibinfo{person}{Mingli Song}.} \bibinfo{year}{2024}\natexlab{c}.
\newblock \showarticletitle{Cola: Cross-city mobility transformer for human trajectory simulation}. In \bibinfo{booktitle}{\emph{Proceedings of the ACM on Web Conference 2024}}. \bibinfo{pages}{3509--3520}.
\newblock


\bibitem[Wang et~al\mbox{.}(2024d)]%
        {wang2024spatiotemporal}
\bibfield{author}{\bibinfo{person}{Yu Wang}, \bibinfo{person}{Tongya Zheng}, \bibinfo{person}{Shunyu Liu}, \bibinfo{person}{Zunlei Feng}, \bibinfo{person}{Kaixuan Chen}, \bibinfo{person}{Yunzhi Hao}, {and} \bibinfo{person}{Mingli Song}.} \bibinfo{year}{2024}\natexlab{d}.
\newblock \showarticletitle{Spatiotemporal-Augmented Graph Neural Networks for Human Mobility Simulation}.
\newblock \bibinfo{journal}{\emph{IEEE Transactions on Knowledge and Data Engineering}} \bibinfo{volume}{36}, \bibinfo{number}{11} (\bibinfo{year}{2024}), \bibinfo{pages}{7074--7086}.
\newblock


\bibitem[Wen et~al\mbox{.}(2022)]%
        {wen2022distributionally}
\bibfield{author}{\bibinfo{person}{Hongyi Wen}, \bibinfo{person}{Xinyang Yi}, \bibinfo{person}{Tiansheng Yao}, \bibinfo{person}{Jiaxi Tang}, \bibinfo{person}{Lichan Hong}, {and} \bibinfo{person}{Ed~H Chi}.} \bibinfo{year}{2022}\natexlab{}.
\newblock \showarticletitle{Distributionally-robust recommendations for improving worst-case user experience}. In \bibinfo{booktitle}{\emph{Proceedings of the ACM Web Conference 2022}}. \bibinfo{pages}{3606--3610}.
\newblock


\bibitem[Wu et~al\mbox{.}(2023b)]%
        {wu2023visual}
\bibfield{author}{\bibinfo{person}{Chenfei Wu}, \bibinfo{person}{Shengming Yin}, \bibinfo{person}{Weizhen Qi}, \bibinfo{person}{Xiaodong Wang}, \bibinfo{person}{Zecheng Tang}, {and} \bibinfo{person}{Nan Duan}.} \bibinfo{year}{2023}\natexlab{b}.
\newblock \showarticletitle{Visual chatgpt: Talking, drawing and editing with visual foundation models}.
\newblock \bibinfo{journal}{\emph{arXiv preprint arXiv:2303.04671}} (\bibinfo{year}{2023}).
\newblock


\bibitem[Wu et~al\mbox{.}(2023a)]%
        {wu2023understanding}
\bibfield{author}{\bibinfo{person}{Junkang Wu}, \bibinfo{person}{Jiawei Chen}, \bibinfo{person}{Jiancan Wu}, \bibinfo{person}{Wentao Shi}, \bibinfo{person}{Xiang Wang}, {and} \bibinfo{person}{Xiangnan He}.} \bibinfo{year}{2023}\natexlab{a}.
\newblock \showarticletitle{Understanding contrastive learning via distributionally robust optimization}.
\newblock \bibinfo{journal}{\emph{Advances in Neural Information Processing Systems}}  \bibinfo{volume}{36} (\bibinfo{year}{2023}), \bibinfo{pages}{23297--23320}.
\newblock


\bibitem[Wu et~al\mbox{.}(2024)]%
        {wu2024survey}
\bibfield{author}{\bibinfo{person}{Likang Wu}, \bibinfo{person}{Zhi Zheng}, \bibinfo{person}{Zhaopeng Qiu}, \bibinfo{person}{Hao Wang}, \bibinfo{person}{Hongchao Gu}, \bibinfo{person}{Tingjia Shen}, \bibinfo{person}{Chuan Qin}, \bibinfo{person}{Chen Zhu}, \bibinfo{person}{Hengshu Zhu}, \bibinfo{person}{Qi Liu}, {et~al\mbox{.}}} \bibinfo{year}{2024}\natexlab{}.
\newblock \showarticletitle{A survey on large language models for recommendation}.
\newblock \bibinfo{journal}{\emph{World Wide Web}} \bibinfo{volume}{27}, \bibinfo{number}{5} (\bibinfo{year}{2024}), \bibinfo{pages}{60}.
\newblock


\bibitem[Yang et~al\mbox{.}(2024)]%
        {yang2024psl}
\bibfield{author}{\bibinfo{person}{Weiqin Yang}, \bibinfo{person}{Jiawei Chen}, \bibinfo{person}{Xin Xin}, \bibinfo{person}{Sheng Zhou}, \bibinfo{person}{Binbin Hu}, \bibinfo{person}{Yan Feng}, \bibinfo{person}{Chun Chen}, {and} \bibinfo{person}{Can Wang}.} \bibinfo{year}{2024}\natexlab{}.
\newblock \showarticletitle{PSL: Rethinking and Improving Softmax Loss from Pairwise Perspective for Recommendation}.
\newblock \bibinfo{journal}{\emph{arXiv preprint arXiv:2411.00163}} (\bibinfo{year}{2024}).
\newblock


\bibitem[Yang et~al\mbox{.}(2025)]%
        {yang2025breaking}
\bibfield{author}{\bibinfo{person}{Weiqin Yang}, \bibinfo{person}{Jiawei Chen}, \bibinfo{person}{Shengjia Zhang}, \bibinfo{person}{Peng Wu}, \bibinfo{person}{Yuegang Sun}, \bibinfo{person}{Yan Feng}, \bibinfo{person}{Chun Chen}, {and} \bibinfo{person}{Can Wang}.} \bibinfo{year}{2025}\natexlab{}.
\newblock \showarticletitle{Breaking the top-k barrier: Advancing top-k ranking metrics optimization in recommender systems}. In \bibinfo{booktitle}{\emph{Proceedings of the 31st ACM SIGKDD Conference on Knowledge Discovery and Data Mining V. 2}}. \bibinfo{pages}{3542--3552}.
\newblock


\bibitem[Yang et~al\mbox{.}(2023)]%
        {yang2023generic}
\bibfield{author}{\bibinfo{person}{Zhengyi Yang}, \bibinfo{person}{Xiangnan He}, \bibinfo{person}{Jizhi Zhang}, \bibinfo{person}{Jiancan Wu}, \bibinfo{person}{Xin Xin}, \bibinfo{person}{Jiawei Chen}, {and} \bibinfo{person}{Xiang Wang}.} \bibinfo{year}{2023}\natexlab{}.
\newblock \showarticletitle{A generic learning framework for sequential recommendation with distribution shifts}. In \bibinfo{booktitle}{\emph{Proceedings of the 46th International ACM SIGIR Conference on Research and Development in Information Retrieval}}. \bibinfo{pages}{331--340}.
\newblock


\bibitem[Zhang et~al\mbox{.}(2024a)]%
        {zhang2024agentcf}
\bibfield{author}{\bibinfo{person}{Junjie Zhang}, \bibinfo{person}{Yupeng Hou}, \bibinfo{person}{Ruobing Xie}, \bibinfo{person}{Wenqi Sun}, \bibinfo{person}{Julian McAuley}, \bibinfo{person}{Wayne~Xin Zhao}, \bibinfo{person}{Leyu Lin}, {and} \bibinfo{person}{Ji-Rong Wen}.} \bibinfo{year}{2024}\natexlab{a}.
\newblock \showarticletitle{Agentcf: Collaborative learning with autonomous language agents for recommender systems}. In \bibinfo{booktitle}{\emph{Proceedings of the ACM Web Conference 2024}}. \bibinfo{pages}{3679--3689}.
\newblock


\bibitem[Zhang et~al\mbox{.}(2025b)]%
        {zhang2025bifair}
\bibfield{author}{\bibinfo{person}{Jiaming Zhang}, \bibinfo{person}{Yuyuan Li}, \bibinfo{person}{Yiqun Xu}, \bibinfo{person}{Li Zhang}, \bibinfo{person}{Xiaohua Feng}, \bibinfo{person}{Zhifei Ren}, {and} \bibinfo{person}{Chaochao Chen}.} \bibinfo{year}{2025}\natexlab{b}.
\newblock \showarticletitle{BiFair: A Fairness-aware Training Framework for LLM-enhanced Recommender Systems via Bi-level Optimization}.
\newblock \bibinfo{journal}{\emph{arXiv preprint arXiv:2507.04294}} (\bibinfo{year}{2025}).
\newblock


\bibitem[Zhang et~al\mbox{.}(2025a)]%
        {zhang2025advancing}
\bibfield{author}{\bibinfo{person}{Shengjia Zhang}, \bibinfo{person}{Jiawei Chen}, \bibinfo{person}{Changdong Li}, \bibinfo{person}{Sheng Zhou}, \bibinfo{person}{Qihao Shi}, \bibinfo{person}{Yan Feng}, \bibinfo{person}{Chun Chen}, {and} \bibinfo{person}{Can Wang}.} \bibinfo{year}{2025}\natexlab{a}.
\newblock \showarticletitle{Advancing Loss Functions in Recommender Systems: A Comparative Study with a R{\'e}nyi Divergence-Based Solution}. In \bibinfo{booktitle}{\emph{Proceedings of the AAAI Conference on Artificial Intelligence}}, Vol.~\bibinfo{volume}{39}. \bibinfo{pages}{13286--13294}.
\newblock


\bibitem[Zhang et~al\mbox{.}(2024b)]%
        {zhang2024causality}
\bibfield{author}{\bibinfo{person}{Yang Zhang}, \bibinfo{person}{Juntao You}, \bibinfo{person}{Yimeng Bai}, \bibinfo{person}{Jizhi Zhang}, \bibinfo{person}{Keqin Bao}, \bibinfo{person}{Wenjie Wang}, {and} \bibinfo{person}{Tat-Seng Chua}.} \bibinfo{year}{2024}\natexlab{b}.
\newblock \showarticletitle{Causality-enhanced behavior sequence modeling in LLMs for personalized recommendation}.
\newblock \bibinfo{journal}{\emph{arXiv preprint arXiv:2410.22809}} (\bibinfo{year}{2024}).
\newblock


\bibitem[Zhao et~al\mbox{.}(2024)]%
        {zhao2024explainability}
\bibfield{author}{\bibinfo{person}{Haiyan Zhao}, \bibinfo{person}{Hanjie Chen}, \bibinfo{person}{Fan Yang}, \bibinfo{person}{Ninghao Liu}, \bibinfo{person}{Huiqi Deng}, \bibinfo{person}{Hengyi Cai}, \bibinfo{person}{Shuaiqiang Wang}, \bibinfo{person}{Dawei Yin}, {and} \bibinfo{person}{Mengnan Du}.} \bibinfo{year}{2024}\natexlab{}.
\newblock \showarticletitle{Explainability for large language models: A survey}.
\newblock \bibinfo{journal}{\emph{ACM Transactions on Intelligent Systems and Technology}} \bibinfo{volume}{15}, \bibinfo{number}{2} (\bibinfo{year}{2024}), \bibinfo{pages}{1--38}.
\newblock


\bibitem[Zhao et~al\mbox{.}(2023)]%
        {zhao2023popularity}
\bibfield{author}{\bibinfo{person}{Jujia Zhao}, \bibinfo{person}{Wenjie Wang}, \bibinfo{person}{Xinyu Lin}, \bibinfo{person}{Leigang Qu}, \bibinfo{person}{Jizhi Zhang}, {and} \bibinfo{person}{Tat-Seng Chua}.} \bibinfo{year}{2023}\natexlab{}.
\newblock \showarticletitle{Popularity-aware distributionally robust optimization for recommendation system}. In \bibinfo{booktitle}{\emph{Proceedings of the 32nd ACM International Conference on Information and Knowledge Management}}. \bibinfo{pages}{4967--4973}.
\newblock


\bibitem[Zheng et~al\mbox{.}(2024b)]%
        {zheng2024adapting}
\bibfield{author}{\bibinfo{person}{Bowen Zheng}, \bibinfo{person}{Yupeng Hou}, \bibinfo{person}{Hongyu Lu}, \bibinfo{person}{Yu Chen}, \bibinfo{person}{Wayne~Xin Zhao}, \bibinfo{person}{Ming Chen}, {and} \bibinfo{person}{Ji-Rong Wen}.} \bibinfo{year}{2024}\natexlab{b}.
\newblock \showarticletitle{Adapting large language models by integrating collaborative semantics for recommendation}. In \bibinfo{booktitle}{\emph{2024 IEEE 40th International Conference on Data Engineering (ICDE)}}. IEEE, \bibinfo{pages}{1435--1448}.
\newblock


\bibitem[Zheng et~al\mbox{.}(2024a)]%
        {zheng2024harnessing}
\bibfield{author}{\bibinfo{person}{Zhi Zheng}, \bibinfo{person}{Wenshuo Chao}, \bibinfo{person}{Zhaopeng Qiu}, \bibinfo{person}{Hengshu Zhu}, {and} \bibinfo{person}{Hui Xiong}.} \bibinfo{year}{2024}\natexlab{a}.
\newblock \showarticletitle{Harnessing large language models for text-rich sequential recommendation}. In \bibinfo{booktitle}{\emph{Proceedings of the ACM Web Conference 2024}}. \bibinfo{pages}{3207--3216}.
\newblock


\bibitem[Zhou et~al\mbox{.}(2024)]%
        {zhou2024explaining}
\bibfield{author}{\bibinfo{person}{Wei Zhou}, \bibinfo{person}{Heike Adel}, \bibinfo{person}{Hendrik Schuff}, {and} \bibinfo{person}{Ngoc~Thang Vu}.} \bibinfo{year}{2024}\natexlab{}.
\newblock \showarticletitle{Explaining pre-trained language models with attribution scores: An analysis in low-resource settings}.
\newblock \bibinfo{journal}{\emph{arXiv preprint arXiv:2403.05338}} (\bibinfo{year}{2024}).
\newblock


\bibitem[Zhu et~al\mbox{.}(2024)]%
        {zhu2024collaborative}
\bibfield{author}{\bibinfo{person}{Yaochen Zhu}, \bibinfo{person}{Liang Wu}, \bibinfo{person}{Qi Guo}, \bibinfo{person}{Liangjie Hong}, {and} \bibinfo{person}{Jundong Li}.} \bibinfo{year}{2024}\natexlab{}.
\newblock \showarticletitle{Collaborative large language model for recommender systems}. In \bibinfo{booktitle}{\emph{Proceedings of the ACM on Web Conference 2024}}. \bibinfo{pages}{3162--3172}.
\newblock


\end{thebibliography}


\appendix

\section{Appendices}

\subsection{Additional Validation of Context Bias Across Prompt Templates and LLMs}
\label{apd:FAA}
\begin{figure}[h]
  \centering
  \includegraphics[width=\linewidth]{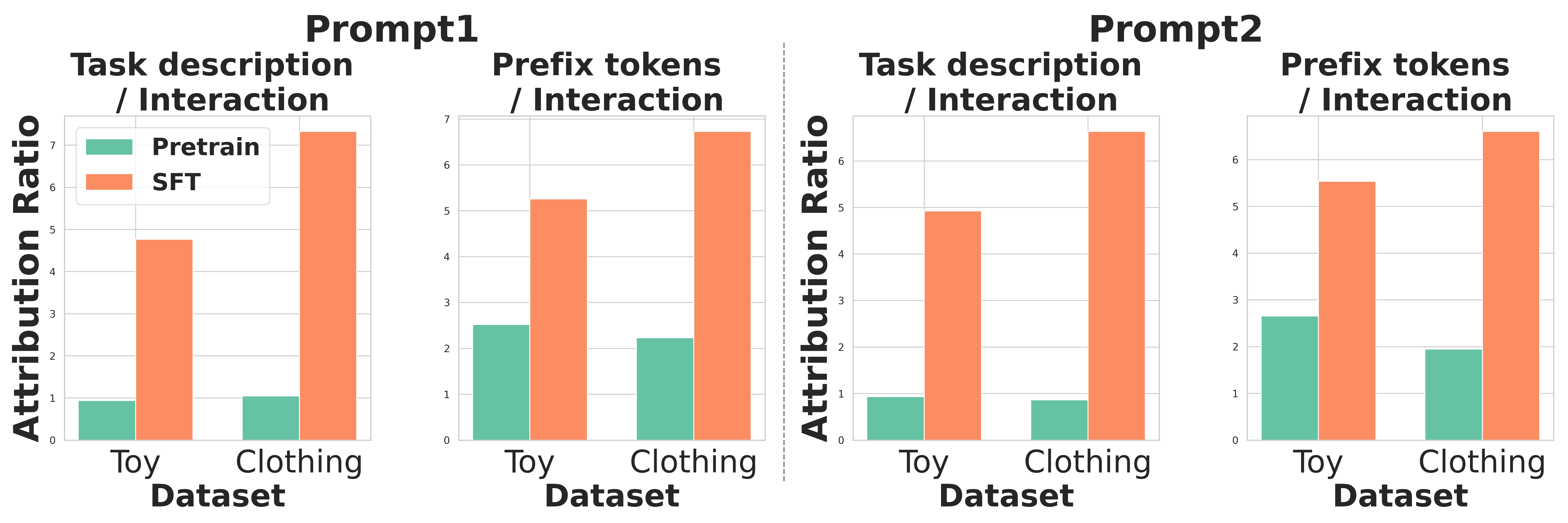}
  \caption{Ratio of attribution values between auxiliary tokens and user-interaction tokens before and after SFT across different prompt templates.}
  \Description{}
  \label{fig:feature_ablation_prompt}
\end{figure}

\begin{figure}[h]
  \centering
  \includegraphics[width=\linewidth]{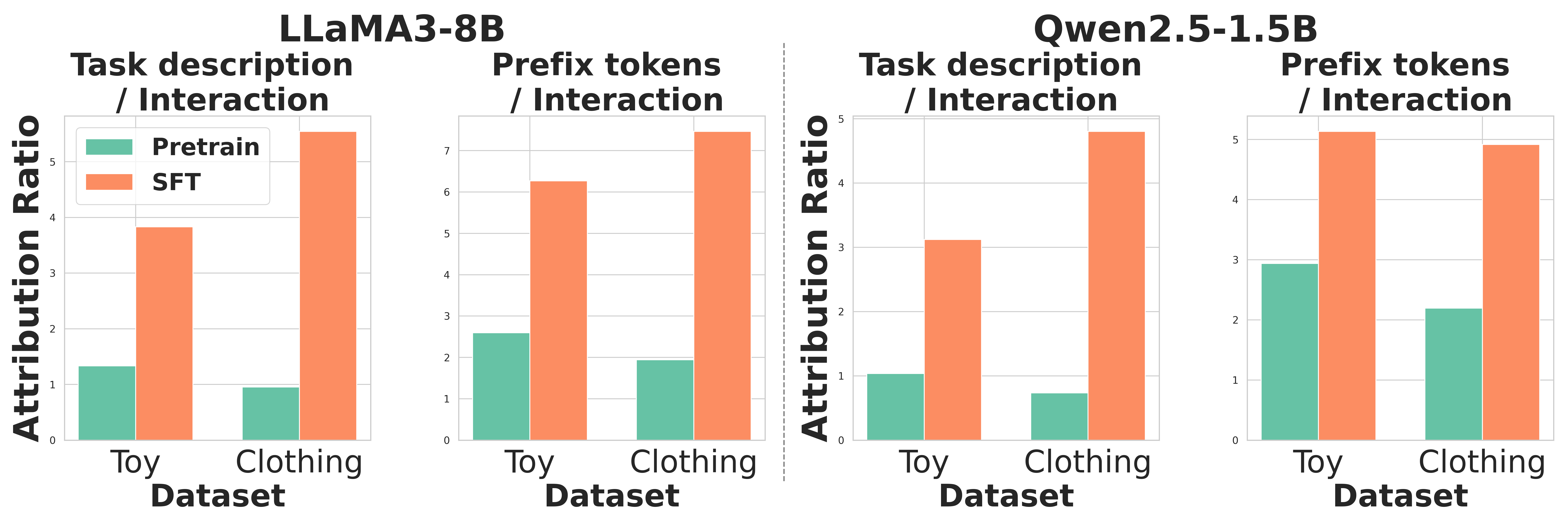}
  \caption{Ratio of attribution values between auxiliary tokens and user-interaction tokens before and after SFT across different LLMs.}
  \Description{}
  \label{fig:feature_ablation_model}
\end{figure}

In this section, to further rule out the potential influence of prompt templates and different LLM backbones on the analysis of context bias, we extend the FAA experiments described in Section~\ref{sec:evidence_bias} to a broader range of prompt templates (see Figure~\ref{fig:feature_ablation_prompt}) and LLM backbones (see Figure~\ref{fig:feature_ablation_model}), beyond those used in the main experiments.
Specifically, for prompt templates, we adopt those proposed in \cite{liao2024llara} and \cite{bao2024decoding}, which we refer to as \textit{Prompt1} and \textit{Prompt2}, respectively. For LLM backbones, we evaluate LLaMA3-8B \cite{dubey2024llama} and Qwen2.5-1.5B \cite{team2024qwen2}.
Across all prompt template and backbone configurations, SFT consistently amplifies the ratio of attribution values between auxiliary tokens and user-interaction tokens. This observation indicates that context bias persistently exists across different prompt templates and LLM backbones. The underlying reason is that the origin of context bias lies in the dataset itself: the co-occurrence rate between auxiliary tokens and target item tokens is significantly higher than that between interaction tokens and target item tokens, as discussed in Section~\ref{sec:origins}.

\subsection{Details of Baselines}
\label{apd:baseline}
The methods compared fall into several categories:
\begin{itemize}[left=5pt]
    \item \textbf{Traditional RS (SASRec \cite{kang2018self}, SASRec++ \cite{lepage2025closing}, DROS \cite{yang2023generic})}: SASRec is a representative sequential recommendation model that employs self-attention mechanisms to effectively capture users’ dynamic interest patterns from historical interaction sequences. Building upon SASRec, SASRec++ introduces an improved training objective by adopting the softmax loss instead of BCE loss used in original SASRec, which leads to more stable optimization and enhanced recommendation performance. DROS incorporates Distributionally Robust Optimization (DRO) into sequential recommendation, aiming to improve model robustness under distributional shifts.
    \item \textbf{LLM-based RS (SFT \cite{bao2025bi}, CFT \cite{zhang2024causality}, MSL \cite{wang2025msl}, LLaRA \cite{liao2024llara}, A-LLM \cite{kim2024large}}): This line of work leverages the strong representation and reasoning capabilities of LLMs for recommendation. SFT applies instruction-tuning strategies with carefully designed templates to adapt LLMs to recommendation tasks. CFT incorporates a causal loss to strengthen the behavior sequence modeling capabilities of LLMs. MSL improves the loss function specifically tailoring it to optimize recommendation-oriented objectives. LLaRA enhances LLM-based recommenders by incorporating embeddings from traditional recommendation models, enabling better exploitation of collaborative filtering signals. A-LLM extends this idea by aligning these collaborative embeddings with their corresponding textual semantics, facilitating more effective integration of structured and unstructured information.
    \item \textbf{Debiasing Methods for LLM-based RS (Reweight \cite{jiang2024item}, SPRec \cite{gao2025sprec}, D3 \cite{bao2024decoding})}: These methods focus on mitigating various biases that arise when applying LLMs to recommendation. Reweight addresses popularity bias by balancing recommendations using pre-calculated item weights. SPRec proposes a popularity-aware negative sampling strategy within Direct Preference Optimization (DPO) \cite{rafailov2024direct} to reduce popularity bias. D3 focuses on mitigating amplification bias during inference by improving the decoding strategy, preventing the model from over-recommending items whose textual representations contain tokens with excessively high generation probabilities.
\end{itemize}

\subsection{Additional Performance Comparison Across Prompt Templates and LLMs}
\label{apd:performance}
In this section, we present additional comparative experiments on the performance of GDRT and SFT that go beyond the prompt templates and LLM backbones used in the main experiments. As summarized in Tables \ref{tab:app_prompt} and \ref{tab:app_model}, GDRT consistently improves recommendation accuracy while achieving substantial gains in fairness across all evaluated configurations, demonstrating strong generalization ability over a broader range of prompts and LLMs.

\begin{table}[h]
\centering
\caption{Performance comparison of SFT and GDRT across different prompt templates. The best result is bolded.}
\label{tab:app_prompt}
\begin{tabular}{@{}l|l|l|cc@{}}
\toprule
Prompt                   & Dataset                   & Method & NDCG@5 & DGU@5  \\ \midrule
\multirow{4}{*}{Prompt1 \cite{liao2024llara}} & \multirow{2}{*}{Toy}      & SFT    & 0.0118 & 0.6549 \\
                         &                           & GDRT   & \textbf{0.0144} & \textbf{0.4616} \\ \cmidrule(l){2-5} 
                         & \multirow{2}{*}{Clothing} & SFT    & 0.0033 & 0.4631 \\
                         &                           & GDRT   & \textbf{0.0052} & \textbf{0.2025} \\ \midrule
\multirow{4}{*}{Prompt2 \cite{bao2024decoding}} & \multirow{2}{*}{Toy}      & SFT    & 0.0116 & 0.5765 \\
                         &                           & GDRT   & \textbf{0.0136} & \textbf{0.3987} \\ \cmidrule(l){2-5} 
                         & \multirow{2}{*}{Clothing} & SFT    & 0.0039 & 0.5231 \\
                         &                           & GDRT   & \textbf{0.0045} & \textbf{0.1893} \\ \bottomrule
\end{tabular}
\end{table}

\begin{table}[h]
\centering
\caption{Performance comparison of SFT and GDRT across different LLMs. The best result is bolded.}
\label{tab:app_model}
\begin{tabular}{@{}l|l|l|cc@{}}
\toprule
LLM                           & \multicolumn{1}{c|}{Dataset} & \multicolumn{1}{c|}{Method} & NDCG@5 & DGU@5  \\ \midrule
\multirow{4}{*}{Llama3-8B}    & \multirow{2}{*}{Toy}         & SFT                         & 0.0151 & 0.6861 \\
                              &                              & GDRT                        & \textbf{0.0173} & \textbf{0.5970} \\ \cmidrule(l){2-5} 
                              & \multirow{2}{*}{Clothing}    & SFT                         & 0.0039 & 0.5801 \\
                              &                              & GDRT                        & \textbf{0.0062} & \textbf{0.2068} \\ \midrule
\multirow{4}{*}{Qwen2.5-1.5B} & \multirow{2}{*}{Toy}         & SFT                         & 0.0098 & 0.4849 \\
                              &                              & GDRT                        & \textbf{0.0117} & \textbf{0.2311} \\ \cmidrule(l){2-5} 
                              & \multirow{2}{*}{Clothing}    & SFT                         & 0.0018 & 0.3538 \\
                              &                              & GDRT                        & \textbf{0.0026} & \textbf{0.0778} \\ \bottomrule
\end{tabular}%
\end{table}

\subsection{The proof of Lemma \ref{lemma1}}
\label{apd:lemma1}
The original formulation of GDRT 
\begin{equation}
\mathcal{L}_{GDRT} = \max_{Q} \sum_{g=1}^{G} Q(g) \mathcal{L}(g),
\quad \text{s.t. } D_{KL}(Q,U) \le \eta,
\end{equation}
can be rewritten in the expectation form as
\begin{equation}
\begin{aligned}
&\max_{Q} \mathbb{E}_{g\sim Q} [\mathcal{L}(g)] \\
s.t. & \mathbb{E}_{g\sim Q}  [\log \frac{Q(g)}{U(g)}] \leq \eta
\end{aligned}
\end{equation}
In the following, we focus on how to eliminate the inner maximization optimization problem and the KL constraint term. Assume $W(g)=Q(g)/U(g)$ and define a convex function $\phi(x)=x \log x-x+1$. Then the divergence $D_{KL}({Q},{U})$ can be written as $\mathbb{E}_{U}[\phi(W)]$. The inner layer maximization optimization problem can be reformulated as follow:
\begin{equation}
\begin{aligned}
& \max_{W}\mathbb{E}_{U}\left[\mathcal{L} W\right] \\
\text{s.t.}\ & \mathbb{E}_{U}[\phi(W)]\leq\eta, \mathbb{E}_{U}[W]=1
\end{aligned}
\end{equation}
As a convex optimization problem, we use the Lagrangian function to solve it:
\begin{equation}
\begin{aligned}
&\min_{\tau\geq0,\beta}\max_{W}\mathbb{E}_{U}\left[\mathcal{L}W\right]-\tau(\mathbb{E}_{U}[\phi(W)]-\eta)+\beta(\mathbb{E}_{U}[W]-1) \\
&=\min_{\tau\geq0,\beta}\left\{\tau\eta-\beta+\tau\max_{W}{\mathbb{E}_{U}}\left[\frac{\mathcal{L}+\beta}{\tau}W-\phi(W)\right]\right\} \\
&=\min_{\tau\geq0,\beta}\left\{\tau\eta-\beta+\tau\mathbb{E}_{U}\left[\max_{W}\left(\frac{\mathcal{L}+\beta}{\tau}W-\phi(W)\right)\right] \right\}
\end{aligned}
\end{equation}
Notice that $\max_{W}\left(\frac{\mathcal{L}+\beta}{\tau}W-\phi(W)\right)=\phi^*(\frac{\mathcal{L}+\beta}{\tau})$ is the convex conjugate function of $\phi(x)$ and we have  $\phi^*(x)=e^x-1$. $W(g)=e^{\frac{\mathcal{L}(g)+\beta}{\tau}}$ when the maximum value is obtained.
\begin{equation}
\begin{aligned}
&\min_{\tau\geq0,\beta}\left\{\tau\eta-\beta+\tau\mathbb{E}_{U}\left[\max_{W}\left(\frac{\mathcal{L}+\beta}{\tau}W-\phi(W)\right)\right] \right\} \\
&=\min_{\tau\geq0,\beta}\left\{\tau\eta-\beta+\tau\mathbb{E}_{U}\left[e^{\frac{\mathcal{L}+\beta}{\tau}}-1\right]\right\} \\
&=\min_{\tau\geq0}\left\{\tau\eta+\tau\log{\mathbb{E}_{U}\left[e^{\frac{\mathcal{L}}{\tau}}\right]}\right\}
\end{aligned}
\end{equation}
where $\beta=-\tau\log{\mathbb{E}_{g \sim U}\left[e^{\frac{\mathcal{L}(g)}{\tau}}\right]}$ and $W(g)=\frac{e^{\frac{\mathcal{L}(g)}{\tau}}}{\mathbb{E}_{g' \sim U}\left[e^{\frac{\mathcal{L}(g')}{\tau}}\right]}$ when the minimum value is obtained. We consider the Lagrange multiplier $\tau$ as a hyperparameter related to the robustness radius $\eta$. Then we can get the unconstrained optimization problem as follows,
\begin{equation}
\mathcal{L}_{GDRT} = \tau\eta+\tau\log\mathbb{E}_{g\sim U} \mathrm{exp}\left(\frac{\mathcal{L}(g)}{\tau}\right)
\end{equation}
where the worst-case distribution
\begin{equation}
    Q^*(g)=U(g)\frac{\mathrm{exp}\left(\mathcal{L}(g)/\tau\right)}{\mathbb{E}_{i\sim U}\left[\mathrm{exp}(\mathcal{L}(i)/\tau)\right]}
\end{equation}
Since $U$ is a uniform distribution
\begin{equation}
    Q^*(g) = \frac{\mathrm{exp}\left(\mathcal{L}(g)/\tau\right)}{\sum_{g'}\left[\mathrm{exp}(\mathcal{L}(g')/\tau)\right]}
\end{equation}
Thus lemma \ref{lemma1} is proven.

\end{document}